\documentclass[twocolumn]{aastex631}
\defcitealias{JK23}{JK23}
\defcitealias{BT08}{BT08}
\defcitealias{SS07}{SS07}

\usepackage{enumitem}
\usepackage{booktabs}
\usepackage{graphicx}
\usepackage{amsmath}
\usepackage[nolist]{acronym}
\usepackage[nameinlink,capitalise,noabbrev]{cleveref}
\crefname{deluxetable}{Table}{Tables}

\newcommand{\kpc}{{\,\rm kpc}}
\newcommand{\Gyr}{{\,\rm Gyr}}
\newcommand{\Myr}{{\,\rm Myr}}
\newcommand{\kms}{{\,\rm km \, s^{-1}}}
\newcommand{\Msun}{{\, M_\odot}}

\begin{document}

\title{Dissecting Bar-Induced Stellar Kinematics in Disk Galaxies: The Bisymmetric Model and Rotation Curve Modifications}

\author[0009-0002-0251-9570]{Seungwon Baek}
\affiliation{Department of Physics \& Astronomy, Seoul National University, Seoul 08826, Republic of Korea}

\author[0000-0003-4625-229X]{Woong-Tae Kim}
\affiliation{Department of Physics \& Astronomy, Seoul National University, Seoul 08826, Republic of Korea}
\affiliation{SNU Astronomy Research Center, Seoul National University, Seoul 08826, Republic of Korea}

\author[0000-0002-7202-4373]{Dajeong Jang}
\affiliation{Department of Physics \& Astronomy, Seoul National University, Seoul 08826, Republic of Korea}

\author[0000-0002-5857-5136]{Taehyun Kim}
\affiliation{Department of Astronomy and Atmospheric Sciences, Kyungpook National University, Daegu 41566, Republic of Korea}

\email{bsw1027@snu.ac.kr, unitree@snu.ac.kr, dj.jang@snu.ac.kr, tkim.astro@gmail.com}


\begin{acronym}
    \acro{LOS}{line-of-sight}
    \acro{PA}{position angle}
    \acro{BPS}{boxy/peanut-shaped}
\end{acronym}

\begin{abstract}
We analyze bars formed in $N$-body simulations to investigate two key aspects of stellar kinematic structure of barred galaxies: the angular distributions of the radial and azimuthal components of stellar velocities, and the impact of bars on rotation curves. We find that stars on bar-supporting $x_1$-like orbits exhibit characteristic sawtooth-like radial velocity patterns and arch-like tangential velocity patterns as a function of azimuth. In contrast, stars on box and disk orbits show little azimuthal variation, effectively smoothing the overall velocity distribution. When averaged over all orbital families, the resulting kinematics are broadly consistent with the bisymmetric model of Spekkens \& Sellwood, with the amplitudes of bar-induced velocity perturbations increasing with bar strength. In addition, bars amplify the radial pressure gradient associated with enhanced random stellar motions, leading to a noticeable reduction in the mean rotational velocity. This effect becomes more pronounced with increasing bar strength, resulting in a shallower rotation curve within the bar region. We discuss our results in the context of the kinematic properties of observed barred galaxies. 
\end{abstract}


\section{Introduction} \label{sec:intro}

Barred galaxies are prevalent in the local universe, with approximately two-thirds of disk galaxies, including those with weak bars, hosting central bars \citep{deVaucouleurs63, deVaucouleurs91, Eskridge00, Knapen00, Whyte02, Laurikainen04, Marinova07, Men07, Aguerri09, Men10, Nair10, Masters11, Men12, Buta15, Diaz16, Diaz19}. While the fraction of barred galaxies decreases at higher redshifts \citep{Abraham99, Sheth08, Cameron10, Melvin14}, bar structures have been detected as far back as $z \sim 3$ \citep{Costantin23, Guo23, Amvrosiadis24, LeConte24, Huertas25, Geron25} and even beyond $z>4$ \citep{Hodge19, Smail23, Tsukui24}. This suggests that bars have played a significant role in the evolutionary history of their host galaxies over a substantial fraction of cosmic time.

In systems without a bar, stellar motions are generally well approximated by circular orbits. The presence of a bar introduces non-axisymmetric gravitational torques that alter both the radial and tangential components of stellar velocities. These perturbations are manifested in enhanced radial velocity dispersions \citep[e.g.,][]{Pinna18, WaloMartin22}, with stronger bars typically producing more pronounced noncircular motions. Indeed, \citet{Kim24} observationally confirmed that stronger bars are associated with more significant noncircular motions in a sample of nearby galaxies. However, since observational data are limited to \ac{LOS} velocities, disentangling the distinct contributions of the bar to the radial and tangential stellar motions remains challenging.

Bar-induced perturbations in stellar motions along the radial and tangential directions give rise to characteristic ``S-shaped" features in observed \ac{LOS} velocity maps \citep[e.g.,][]{Peterson78, Bosma78, Huntley78, Peterson80, Pence81, Kormendy83, Pence84, Weliachew88, Fathi05, Stark18}. However, decomposing the \ac{LOS} velocity field into radial and tangential components is a nontrivial task. To address this, \citet{SS07} proposed the bisymmetric model, in which the radial and tangential velocities are described by double-angle sinusoidal functions:
\begin{align}\label{eq:bisymmetric0}
\begin{split}
V_T(R,\phi) &= V_{T,0}(R) - V_{T,2}(R) \cos [2(\phi-\phi_b)], \\ 
V_R(R,\phi) &=   -V_{R,2}(R) \sin [2(\phi-\phi_b)],
\end{split}
\end{align}
where $V_{T,0}(R)$ is the circular velocity in the absence of the bar, $V_{T,2}(R)$ and $V_{R,2}(R)$ denote the amplitudes of the bar-induced tangential and radial velocity perturbations, respectively, and $\phi_b$ is the \ac{PA} of the bar’s semi-major axis.
\Cref{eq:bisymmetric0} indicates that the radial and tangential velocity perturbations are offset by $45^\circ$ in azimuthal phase.

The bisymmetric model has been widely applied in kinematic analyses of high-resolution observational data for barred galaxies \citep[e.g.,][]{Hallenbeck14, Holmes15, LopezCoba22, Zanger24, Hogarth24, Kim24, LopezCoba24a, LopezCoba24b}, using various tools such as {\tt DiskFit} \citep{SS15}, {\tt XookSuut} \citep{LopezCoba24c}, and {\tt Nirvana} \citep{Zanger24}. For example, \citet{LopezCoba22} modeled the stellar and gaseous velocity fields of barred galaxies using data from VLT/MUSE, finding that bars are the primary drivers of the noncircular motions observed in both stars and gas. Notably, such motions enabled the detection of a weak bar in NGC 1087 \citep{LopezCoba24b}. Through statistical analysis, \citet{Hogarth24} further demonstrated that barred galaxies are more likely to exhibit radial gas motions than their unbarred counterparts, as revealed by ALMA CO velocity maps.

Although \cref{eq:bisymmetric0} has been adopted in numerous observational kinematic analyses of barred galaxies, its validity has yet to be tested in numerical simulations. Theoretically, the bisymmetric velocity pattern emerges from the dominance of the $m=2$ Fourier mode induced by the bar. In this configuration, tangential velocities reach their maxima and minima along the bar's minor and major axes, respectively, while the radial component is phase-shifted by 45$^\circ$ relative to the tangential component \citep{SW93,SS07}. Since sinusoidal deviations are usually appropriate only for low-amplitude perturbations, it remains uncertain how accurately the bisymmetric model captures stellar motions in galaxies with strong bars. Observations indicate that the kinematic \ac{PA} of noncircular motions closely aligns with the photometric \ac{PA} of the bar when analyzed using the bisymmetric model \citep{LopezCoba22, Zanger24}. This suggests that \cref{eq:bisymmetric0} provides a reasonable approximation, at least in an averaged sense. Recently, \citet{Ghosh25} demonstrated that a quadrupole pattern in the distribution of the mean radial velocity, as described by \cref{eq:bisymmetric0}, is strongly correlated with the properties of bars in galaxies formed within cosmological simulations.

In this paper, we examine the extent to which \cref{eq:bisymmetric0} captures the stellar kinematics of barred galaxies by analyzing $N$-body simulation data presented in \citet{JK23}. The disk is divided into concentric annuli of varying radii, within which we compute the mean and dispersion of radial and tangential velocities. These mean velocities are then compared with the predictions of \cref{eq:bisymmetric0} to evaluate its applicability. We also classify stellar orbits based on their morphological characteristics and angular momenta and use this classification to elucidate the physical basis for the bisymmetric model as a plausible first-order approximation for the angular variation of stellar motions.

In addition to inducing angular variations in the perturbed velocities, the presence of a bar also imprints a distinct signature on the rotation curve of the galaxy. Using numerical simulations, \citet{Bureau05} found that barred galaxies viewed along the semimajor axis of the bar exhibit a pronounced ``double-hump" structure in the \ac{LOS} velocity distributions, characterized by a steep rise at small radii, a dip or plateau at intermediate radii, and a gradual increase toward a flat outer region. This distinctive feature has also been observed in rotation curves of barred galaxies and is widely regarded as evidence for the presence of a bar \citep[e.g.,][]{Peterson80, Marquez02, Chung04, Seidel15, Molaeinezhad16, Saburova17}.
For instance, NGC 1300 has been observed to exhibit a velocity turnover in its rotation curve \citep{Peterson80}. \citet{Chung04} investigated edge-on disk galaxies with \ac{BPS} bulges and found that nearly $90\%$ of their sample displayed the double-hump feature in their rotation curves. This supports the notion that \ac{BPS} bulges are thickened bars viewed edge-on. \citet{Seidel15} also reported that around $60\%$ of barred galaxies display the double-hump feature in their velocity profiles.

\begin{figure}[t]
\plotone{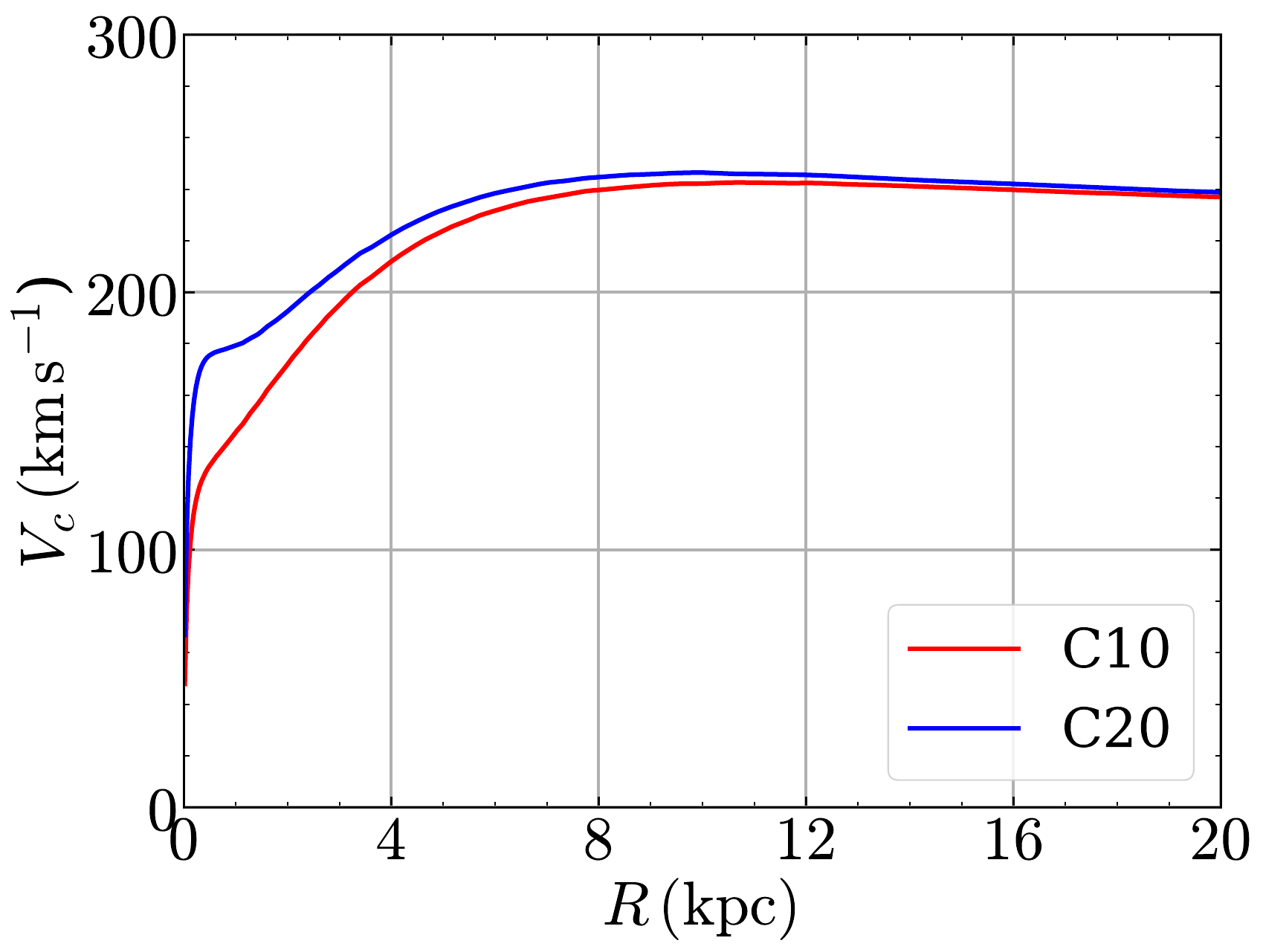}
\caption{Radial distributions of circular velocity, $V_{c}$, for models {\tt C10} (red) and {\tt C20} (blue), calculated from their respective gravitational potentials. Model {\tt C20} exhibits higher $V_{c}$ at small radii as it possesses a more massive bulge.
\label{fig:vc_initial}}
\end{figure}

Although double-hump features are observed in the radial profiles of \ac{LOS} velocities, it remains uncertain whether similar structures are present in the true rotational velocities, i.e., $V_{T,0}$ in \cref{eq:bisymmetric0}, of barred galaxies. If such features do exist, what is their physical origin? \citet{Bureau05} argued that the feature is caused by the (mostly inner) $x_1$ orbits. More recently, \cite{Pejch23} demonstrated that a circumnuclear stellar disk, with a mass two orders of magnitude smaller than that of the main stellar disk, can produce a prominent double-hump structure in gaseous rotation curves. In addition, \citet{Liu25} attributed the ``dip'' feature to the perpendicular alignment between gas flows in the nuclear ring and the bar, and noted that such features are frequently observed in the PHANGS-ALMA sample of barred galaxies. In this paper, we analyze $N$-body simulation data of barred galaxies and demonstrate that bar-induced random motions are responsible for the reduction of the rotation curve in the bar region, giving rise to a double-hump-like feature. We will also show that the reduction in rotational velocity becomes more pronounced with increasing bar strength, leading to a shallower rotation curve within the bar region.

The paper is organized as follows. In \autoref{sec:method}, we describe the model galaxies used in our analysis and the method of orbit classification. In \autoref{sec:bisymmetric}, we examine the angular distributions of radial and tangential velocities, beginning with theoretical $x_1$ orbits and extending to particles in $N$-body bars. We also investigate the relationship between the amplitude of bar-induced velocity perturbations and bar strength. In \autoref{sec:rotation}, we analyze how a bar reduces the rotation curve within the bar region. In \autoref{sec:discussion}, we place our findings in the context of observations of barred galaxies. Finally, we summarize our results in \autoref{sec:conclusion}.

\section{Galaxy Model and Orbit Classification}\label{sec:method}

To investigate the stellar kinematics in barred galaxies, we adopt two Milky Way-like galaxy models from \citet{JK23}. In this section, we briefly describe these models and outline the methods of orbit integration and frequency analysis employed for orbit classification.

\subsection{Galaxy Models}\label{sec:model}

\begin{figure}[t]
\plotone{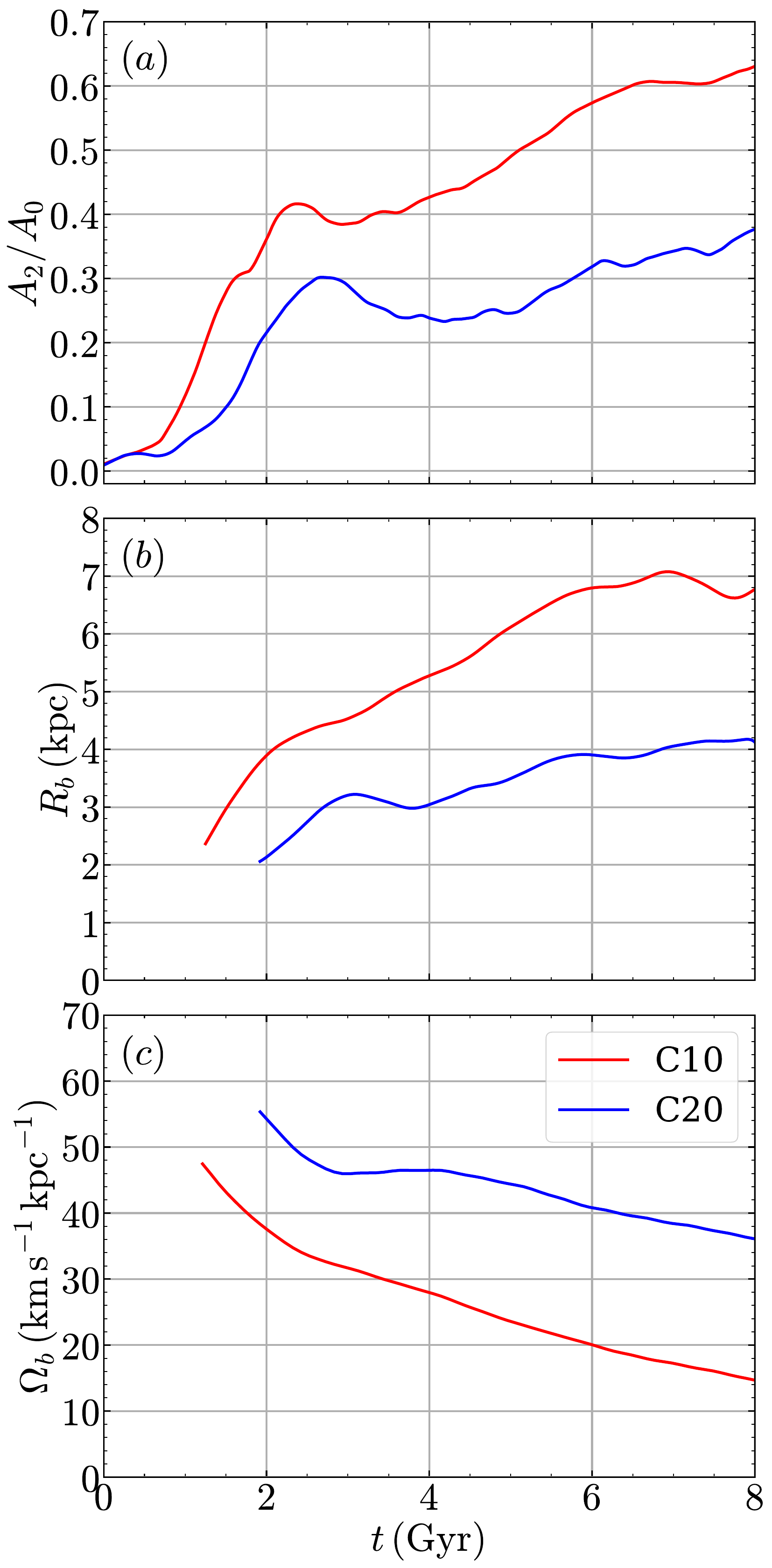}
\caption{Time evolution of (a) the bar strength $A_{2}/A_{0}$, (b) the bar length $R_b$, and (c) the bar pattern speed $\Omega_b$ for models {\tt C10} (red) and {\tt C20} (blue).}
\label{fig:bar_properties}
\end{figure}

\begin{figure*}[t]
\plotone{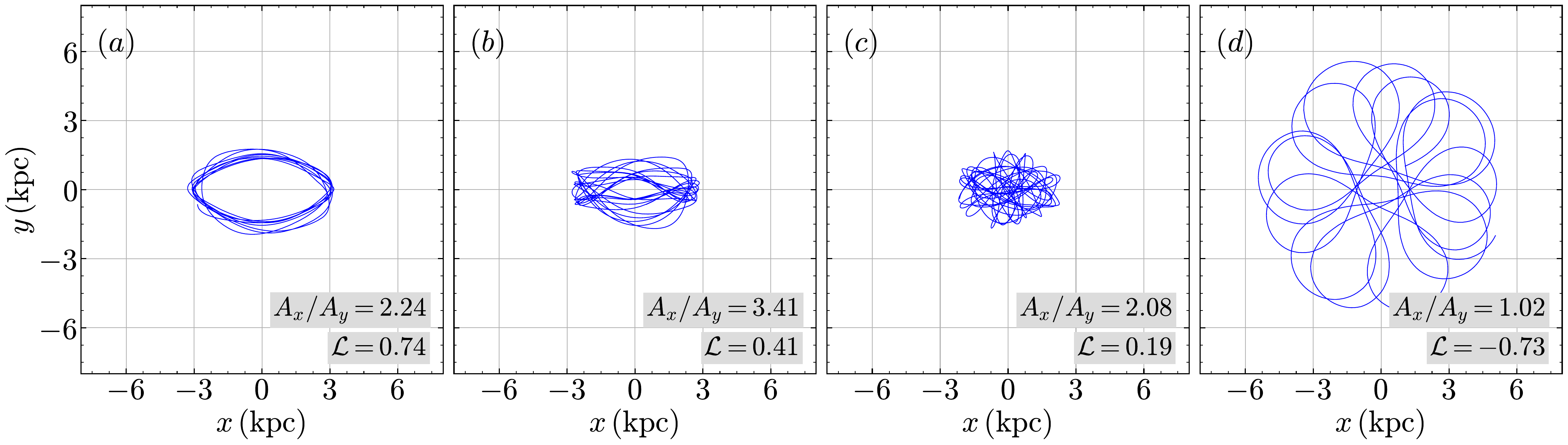}
\caption{Representative face-on views of ($a$, $b$) $x_1$-like orbits, ($c$) a box orbit, and ($d$) a disk orbit in model {\tt C20} at $t=6\Gyr$.
\label{fig:orbits_shape}}
\end{figure*}

From \citet{JK23}, we adopt models \texttt{C10} and \texttt{C20}. These models closely resemble the Milky Way in terms of the physical properties of their bars \citep[e.g.,][]{Shen10,Bland-Hawthorn16, Helmi20}. Both models consist of a stellar disk, a central bulge, a dark halo, and a central supermassive black hole with a mass of $M_\text{BH}=4\times 10^{6} \Msun$. In both models, the disk follows a standard exponential–sech$^2$ vertical density profile, with a total mass of $M_d = 5 \times 10^{10}\Msun$, a scale radius of $R_d = 3\kpc$, and a scale height of $z_d = 0.3\kpc$. 
The dark matter halo is modeled with the \citet{Hernquist90} profile, having a mass of $M_h = 1.3 \times 10^{12}\Msun$ and a scale radius of $a_h = 30\kpc$. The two models differ in the mass of the bulge, which also follows a Hernquist profile, with masses of $M_b = 0.1\,M_d$ and $0.2\,M_d$ for models \texttt{C10} and \texttt{C20}, respectively, and a fixed scale radius of $a_b = 0.4\kpc$. 
\cref{fig:vc_initial} plots the radial distributions of the circular velocity $V_{c}=(R\partial \Phi/\partial R)^{1/2}$ for both models at $t=0$, calculated from their respective gravitational potentials $\Phi$. Model {\tt C20} has higher circular velocities at small radii reflecting the influence of its more massive bulge.

\begin{figure*}[t!]
\plotone{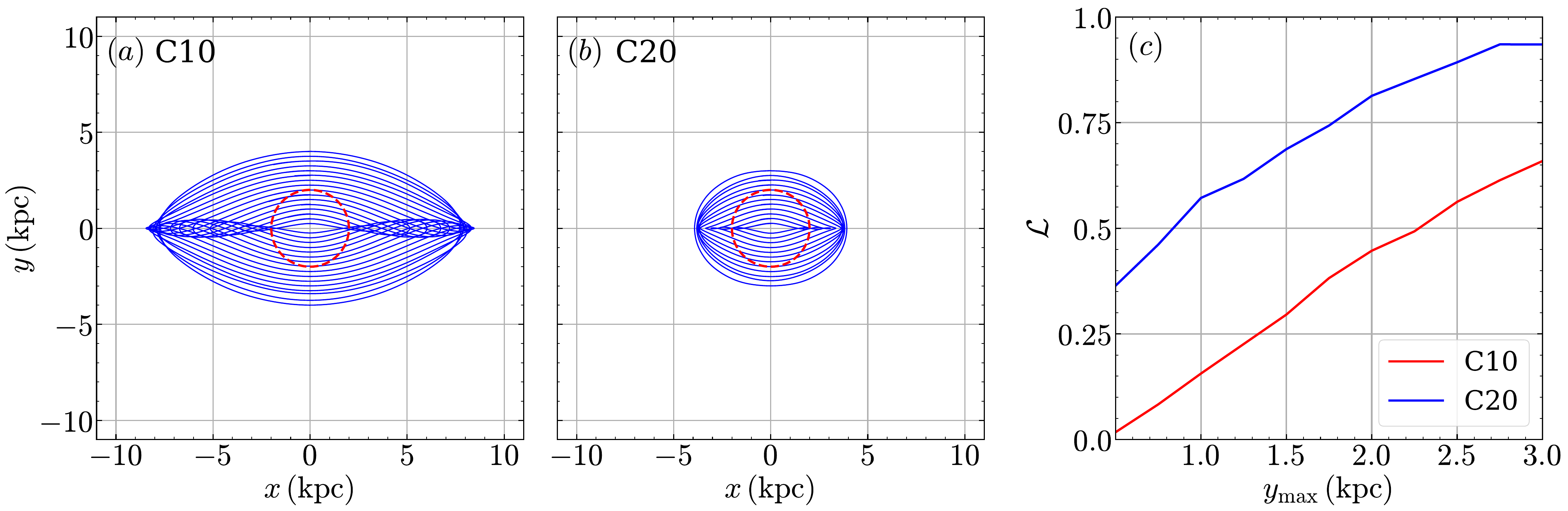}
\caption{Shapes of $x_1$ orbits in the $x$--$y$ plane for the bar in models ($a$) {\tt C10} and ($b$) {\tt C20} at $t=8\Gyr$. The dashed circles indicate a radius of $R = 2\kpc$, along which the radial and tangential velocities are measured, as shown in \cref{fig:frozen_x1_velocity}. ($c$) Dependence of the dimensionless angular momentum $\mathcal{L}$, as defined in \cref{eq:L}, on the maximum excursion $y_\text{max}$ of the $x_1$ orbits along the semiminor axis.}
\label{fig:frozen_x1_face_on}
\end{figure*}

The model galaxies are constructed with $N_h = 2.6 \times 10^7$ halo particles, $N_d = 1.0 \times 10^6$ disk particles, and $N_b = 1.0 \times 10^5$ and $2.0 \times 10^5$ bulge particles for models \texttt{C10} and \texttt{C20}, respectively.
The mass of a single particle is fixed to $5\times10^{4} \Msun$, regardless of its type.
The models are evolved for $t = 10\Gyr$ using a public version of {\tt Gadget-4} \citep{Springel21}.
We refer the reader to \citet{JK23} for a detailed description of the models and their evolution.

Initial perturbations in the random particle distributions grow through successive swing amplifications combined with multiple feedback loops, eventually leading to bar formation. \cref{fig:bar_properties}($a$) plots the temporal evolution of the bar strength, $A_2/A_0$, defined as the maximum Fourier amplitude of the $m=2$ mode across the radius. We define a non-axisymmetric structure as a bar when $A_2/A_0 \ge 0.2$ \citep{Algorry17}.
\cref{fig:bar_properties}($b$) plots the time evolution of the bar length, $R_b$, defined as the radial range over which the bar \ac{PA} varies by less than 0.1 radians ($=5.7^{\circ}$).
\cref{fig:bar_properties}($c$) presents the temporal evolution of the bar pattern speed, $\Omega_b$, defined as the time derivative of the bar \ac{PA}.
Both models form a bar at $t \lesssim 2 \Gyr$, although the more massive bulge in model {\tt C20} delays bar formation. Model {\tt C10} develops a bar that is stronger, longer, and slower than that of model {\tt C20}.

\subsection{Orbit Classification} \label{sec:orbit_integration}

To analyze the orbital structure of disk particles, we utilize data dumps saved every $1\Myr$ from the $N$-body simulations. We track the positions $\mathbf{x}=(x, \, y, \, z)$ and velocities $\mathbf{v}=(v_{x}, \, v_{y},\, v_{z})$ of particles for a $1\,\mathrm{Gyr}$ interval centered at $t=6\Gyr$ (i.e., $5.5\Gyr \le t \le 6.5 \Gyr$), in a reference frame where the bar is aligned with the $x$-axis \citep[e.g.,][]{Gajda16, Smirnov21, Valencia23}. This generates time series data containing 1001 data points for each particle.
With an integration time of $1\Gyr$, which far exceeds the typical orbital period of $\sim 100\Myr$ at half the bar length, the orbits complete numerous cycles, ensuring sufficient sampling for our analysis.
In contrast to traditional approaches that integrate orbits within a frozen potential \citep[e.g.,][]{Voglis07, Valluri16, Chaves17, Patsis18}, our method directly incorporates the time-dependent potentials from $N$-body simulations, ensuring that the traced orbits more accurately represent the actual motion of the particles.

The method of spectral dynamics, first introduced by \citet{Binney82, Binney84}, has been adopted to study the orbital structure of $N$-body systems. This method was later improved by \citet{Laskar90, Laskar93} and \citet{Papaphilippou96, Papaphilippou98} to calculate fundamental frequencies. \citet{Valluri98} and \citet{Valluri10, Valluri16} then developed automated schemes to classify orbits based on these frequencies. In this work, we adopt the method of \citet{Parul20} to improve frequency resolution by zero-padding the time-domain signal for a duration of $9\Gyr \mathrm{s}$ before applying the Fourier transform, which effectively increases the density of frequency bins without altering the original signal content. By performing a discrete Fourier transform, we identify the peak frequencies of stellar motion in the $x$- and $y$-directions, and corresponding peak amplitudes $A_x$ and $A_y$, with a frequency resolution of $\Delta f = 10^{-4} \Myr^{-1}$. We note that the peak frequencies we identify are not influenced by the temporal variations in the bar pattern speed, as it decreases by less than $20\%$ over $1\Gyr$ in our models (see \cref{fig:bar_properties}($c$)).

Another factor influencing the orbit type is the angular momentum, $L_{z} = xv_{y} - yv_{x}$. We introduce a time-averaged, dimensionless angular momentum defined by
\begin{equation}\label{eq:L}
\mathcal{L} \equiv  \frac{\int_0^T L_z dt}{T \max (L_z)},
\end{equation} 
where $T=1\Gyr$ is the duration of the orbit integration. Note that $\mathcal{L}$ quantifies the net prograde rotation over the course of the entire orbit \citep{Valluri16}. A prograde orbit with constant angular momentum yields $\mathcal{L}=1$, while significant temporal variation in $L_z$ results in a lower value of $\mathcal{L}$.

Since a bar in $N$-body simulations is non-steady, with its strength, length, and pattern speed varying over time, particles in the disk do not follow $x_1$ orbits exactly. Nevertheless, a substantial fraction of disk particles exhibit $x_1$-like orbits that support the bar. To determine whether a particle follows a bar-supporting orbit or not, we classify orbits into three categories -- $x_1$-like orbits, box orbits, and disk orbits -- based on the axis ratio $A_x/A_y$ and the dimensionless angular momentum $\mathcal{L}$. A particle is identified as following a bar-supporting orbit if $A_x / A_y \geq 1.5$, indicating intrinsic elongation along the bar \citep{Lokas25}. Among these, particles with $\mathcal{L} \geq 0.25$ are classified as $x_1$-like orbits, while those with $\mathcal{L} < 0.25$ are designated as box orbits, as the latter typically exhibit lower $\mathcal{L}$ values due to prolonged retrograde motion \citep{Valluri16}. In contrast, particles with $A_x / A_y < 1.5$ are categorized as disk orbits.\footnote{The threshold values for $A_x/A_y$ and $\mathcal{L}$ are empirically chosen to ensure clear separation in the angular variation of velocities. Tests with $\pm10\%$ variations in these thresholds show that the main trends and conclusions remain unchanged, demonstrating the robustness of our classification.}

\cref{fig:orbits_shape} plots representative face-on views of two $x_1$-like orbits, a box orbit, and a disk orbit in model {\tt C20} at $t=6\Gyr$. Clearly, $x_1$-like orbits are elongated along the bar, with higher values of $\mathcal{L}$ corresponding to less boxy shapes \citep[e.g., fat $x_1$ orbits in][]{SS04}. Although the box orbits undergo a substantial phase of retrograde motion, they remain confined within the bar region. The disk orbits, by comparison, are not elongated along the bar and therefore do not contribute to supporting the bar structure. The fractions of particles on these types vary over radius and time, but typically $0.75:0.15:0.10$ for the $x_1$-like, box, and disk orbits in model {\tt C10}, respectively, and $0.7: 0.1:0.2$ in model {\tt C20} when measured at $R=0.5R_b$.
It should be noted that our orbit classification scheme is based on orbits projected onto the $x$–$y$ plane, and does not account for vertical motions, as they are not relevant to the present study.

\section{Angular Distribution of Stellar Velocities}\label{sec:bisymmetric}

In this section, we analyze the azimuthal distribution of stellar velocities in our models. We also investigate the relationship between the amplitudes of bar-induced velocity perturbations and bar strength.

\subsection{$x_1$ Orbits}\label{sec:bisymmetric_x1}

It is well known that stars in a barred galaxy predominantly follow $x_1$ orbits \citep{BT08}.
The $x_1$ orbits are characterized by two radial oscillations during one complete prograde revolution \citep{Voglis07}. \cref{fig:frozen_x1_face_on} presents examples of $x_1$ orbits for the bars in models {\tt C10} and {\tt C20} at $t=8\Gyr$, shown in the reference frame where the bar is aligned with the $x$-axis. Also shown is the relation between the maximum extent in the $y$-direction, $y_\mathrm{max}$, and the time-averaged dimensionless angular momentum $\mathcal{L}$.
In general, $\mathcal{L}$ is higher for $x_1$ orbits with larger $y_{\mathrm{max}}$, where the oscillation in $L_z$ is relatively mild. In contrast, some $x_1$ orbits exhibit self-intersections that form prominent ``ears". Because the motion within these ears is partially retrograde, $\mathcal{L}$ can approach zero for orbits with well-developed ears, as illustrated in \cref{fig:frozen_x1_face_on}$(c)$.

\begin{figure}[t!]
\plotone{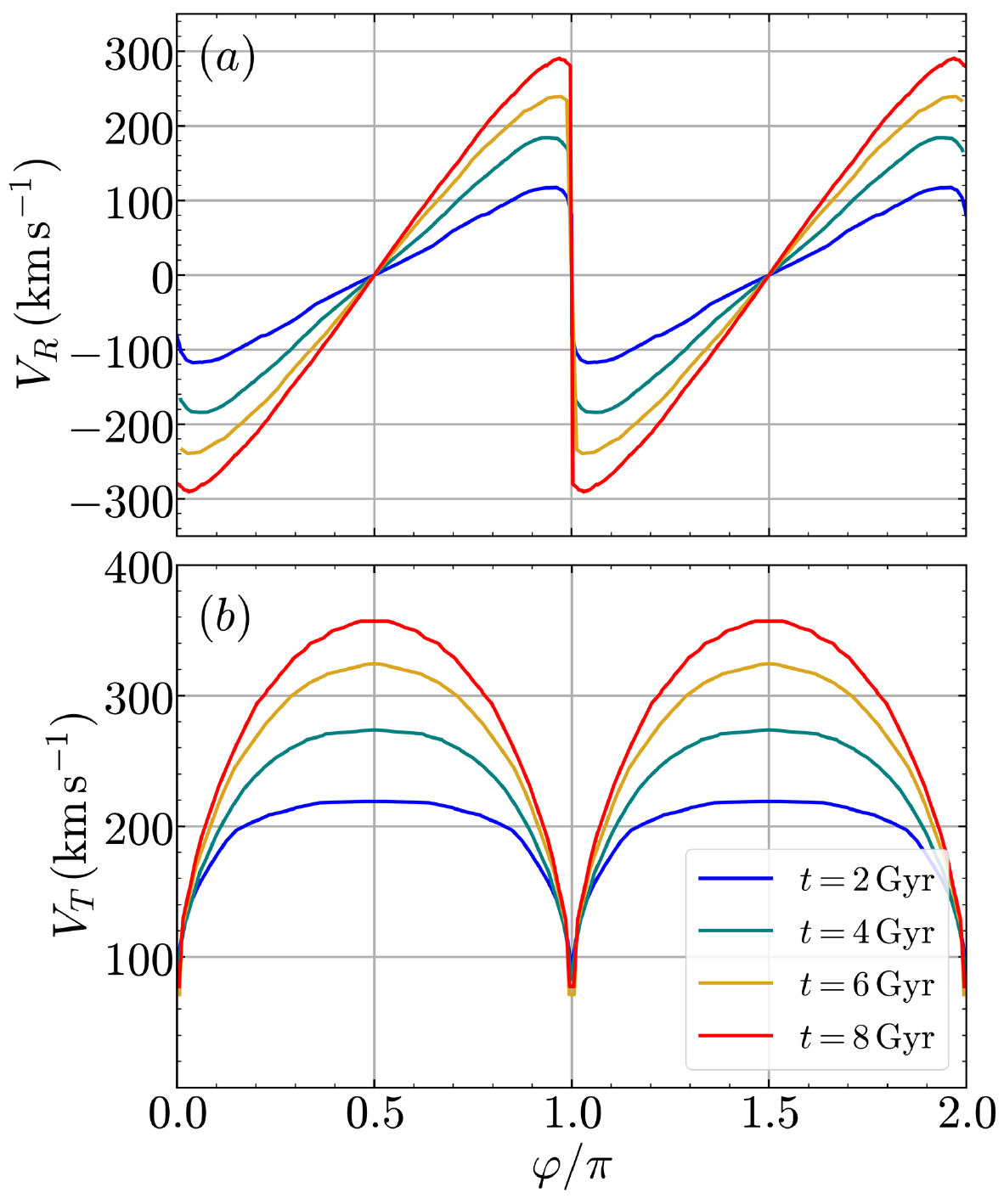}
\caption{Distributions of ($a$) the radial velocity $V_R$ and ($b$) tangential velocity $V_T$ of the $x_1$ orbits in model {\tt C10}, measured at $R=2\kpc$, as functions of the azimuthal angle $\varphi\equiv \phi-\phi_b$ relative to the semimajor axis of the bar. Note that $V_R$ and $V_T$ exhibit sawtooth-like and arch-like patterns, respectively, with higher amplitudes for a stronger bar. }
\label{fig:frozen_x1_velocity}
\end{figure}

\cref{fig:frozen_x1_velocity} plots the angular distribution of the radial and tangential velocities of particles, measured at $R=2\kpc$, moving on $x_1$ orbits. These velocities are calculated at the points of interaction between the $x_1$ orbits and the dashed circles as shown in \cref{fig:frozen_x1_face_on}. The radial velocity exhibits a sawtooth-like pattern such that $V_R$ experiences a rapid drop at $\varphi \equiv \phi-\phi_b =0, \, \pi,$ and $2\pi$, while increasing almost linearly between these points. In contrast, the tangential velocity follows an arch-like pattern rooted at $\varphi = 0, \, \pi,$ and $2\pi$, peaking between these points, with the highest values occurring approximately at $\varphi=\pi/2$ and $3\pi/2$.

\begin{figure}[t!]
\plotone{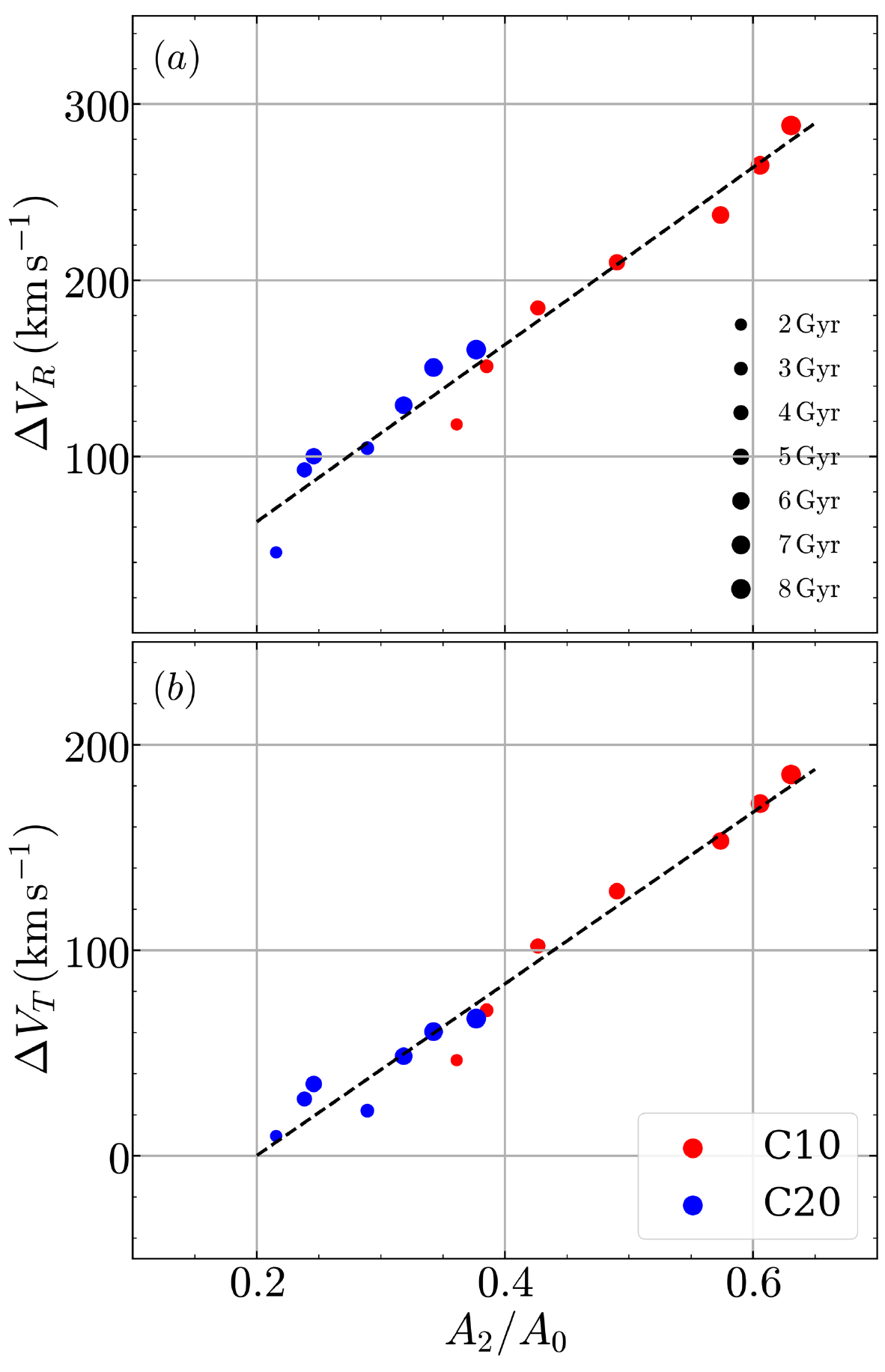}
\caption{Amplitudes of the bar-induced perturbations in ($a$) radial velocity $\Delta V_{R}$ and ($b$) tangential velocity $\Delta V_{T}$ of stellar motions on $x_1$ orbits, measured at $R=2 \kpc$, for models {\tt C10} (red) and {\tt C20} (blue) as functions of the bar strength $A_{2}/A_{0}$. The marker sizes represent simulation times. The black dashed lines are the best-fit relations shown in \cref{eq:x1fits}.} 
\label{fig:x1_correlation}
\end{figure}

These characteristic sawtooth-like and arch-like patterns arise naturally from the elongation of $x_1$ orbits parallel to the bar. Suppose a situation where one moves along the circle with $R = 2\kpc$ starting from the semiminor axis of the bar in the counterclockwise direction. At $\varphi = \pi/2$, the stellar motion is entirely tangential, with $V_R = 0$. As $\varphi$ increases, the radial velocity rises while the tangential velocity declines. At $\varphi = \pi$, $V_R$ undergoes an abrupt sign reversal, while $V_T$ reaches its minimum. Beyond this point, as $\varphi$ increases toward $\varphi = 3\pi/2$, both $V_R$ and $V_T$ increase, completing the half-period variation of the velocities along the circle.

\begin{figure*}[t]
\plotone{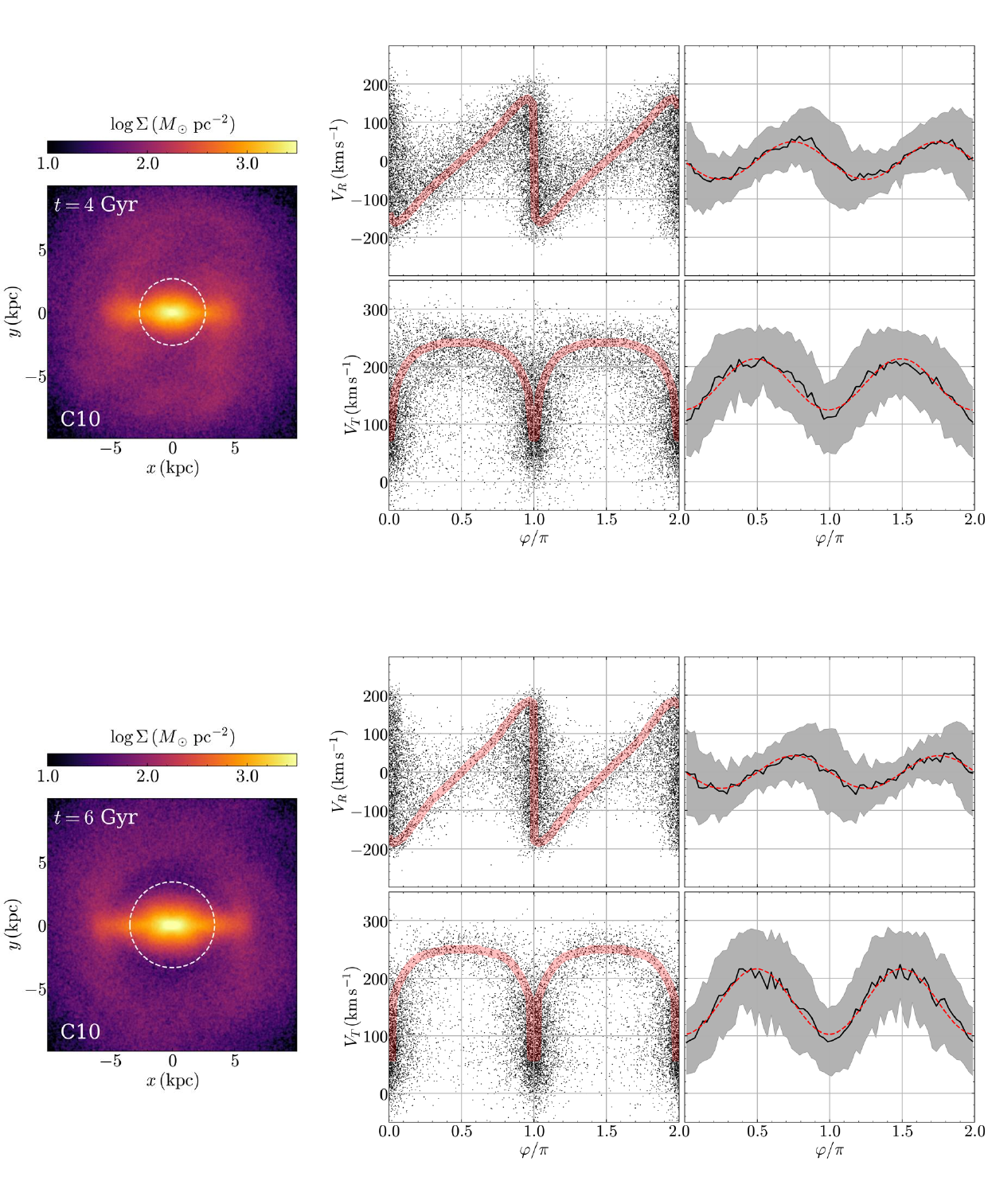}
\caption{Left: Face-on views of the disk surface density $\Sigma$, with a dashed circle indicating a radius of $R=0.5R_b$. Middle: Angular distributions of the radial and tangential velocities of disk particles at $R=0.5R_b$, plotted as dots. Thick red lines indicate the velocities of particles on $x_1$ orbits. Right: Boxcar-averaged mean (black solid line) and standard deviations (shaded area) of the stellar velocities shown in the middle panels. The red dashed lines plot the bisymmetric model, \cref{eq:bisymmetric0}, with appropriately chosen coefficients. The upper and lower panels correspond to model {\tt C10} at $t=4$ and $t=6\Gyr$, respectively.
}
\label{fig:bisymmetric_C10}
\end{figure*}

\begin{figure*}[t]
\plotone{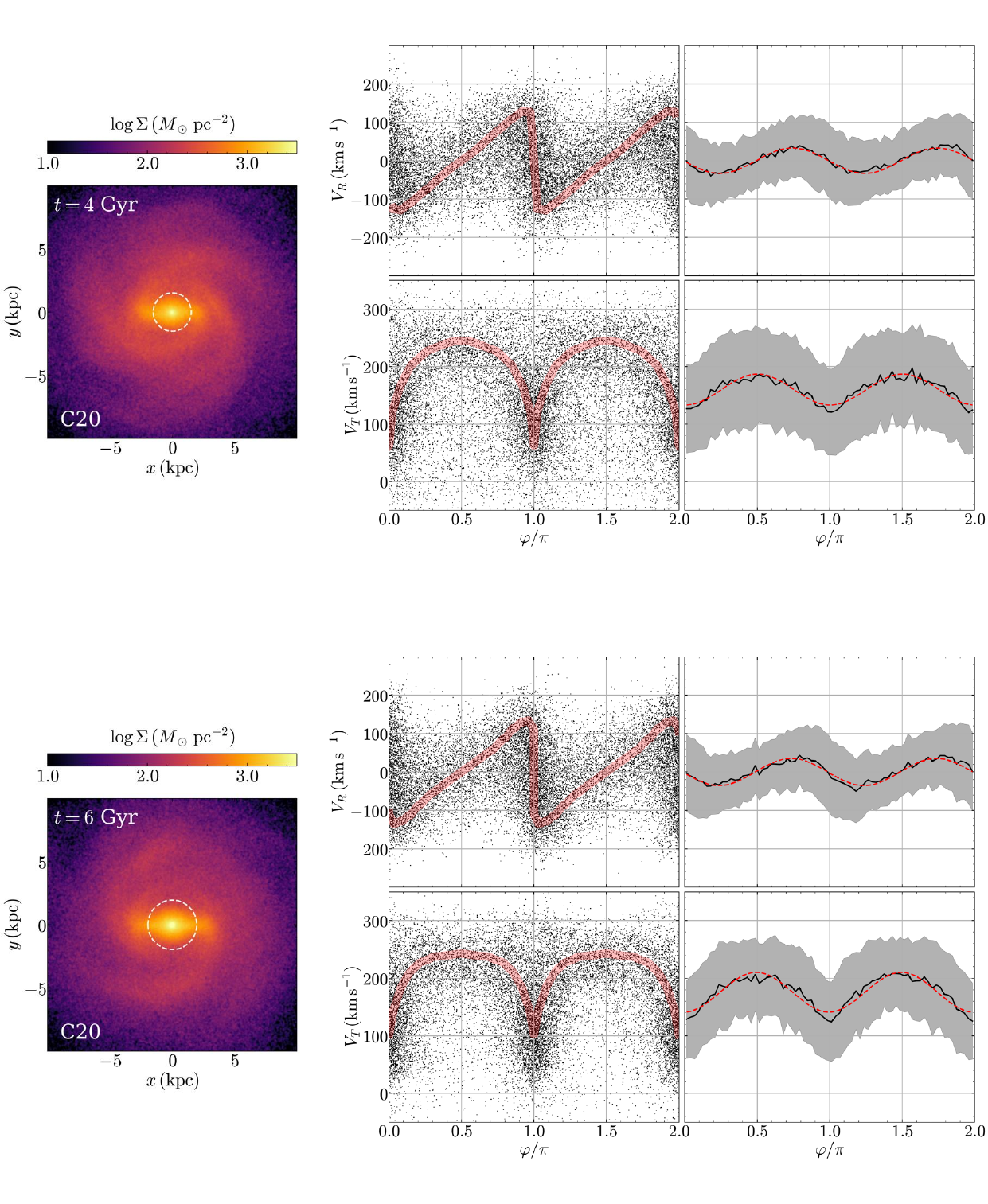}
\caption{Same as \cref{fig:bisymmetric_C10}, but for model {\tt C20}.}
\label{fig:bisymmetric_C20}
\end{figure*}

\cref{fig:bar_properties}($a$) and \cref{fig:frozen_x1_velocity} indicate that the peak values of $V_R$ and $V_T$ increase with bar strength. To quantify the amplitude of the velocity perturbations induced by the bar, we define $\Delta V_R\equiv \max (V_R)$ and $\Delta V_T \equiv \max(V_T- V_c)$ at $R=2\kpc$.  \cref{fig:x1_correlation} plots $\Delta V_R$ and $\Delta V_T$ as functions of $A_2/A_0$ for models \texttt{C10} (red) and \texttt{C20} (blue) over the time interval $t=2$--$8\Gyr$. The velocity perturbations for particles on the $x_1$ orbits clearly exhibit a strong correlation with bar strength. The dashed lines are the best fits 
\begin{equation}\label{eq:x1fits}
\begin{split}
\Delta V_R &= 499 (A_2/A_0) -36 \kms,\\
\Delta V_T &= 418 (A_2/A_0) -83 \kms,
\end{split}
\end{equation}
for $0.2\leq A_2/A_0\leq 0.65$. This is qualitatively consistent with previous findings that bars are the primary drivers of noncircular stellar motions in galactic disks \citep[e.g.,][]{LopezCoba22, Kim24}.
Note that these relations should not be extrapolated to values of $A_2/A_0 < 0.2$, which lie below the bar-formation threshold and thus fall outside the regime of physical relevance.

\begin{figure*}[t]
\plotone{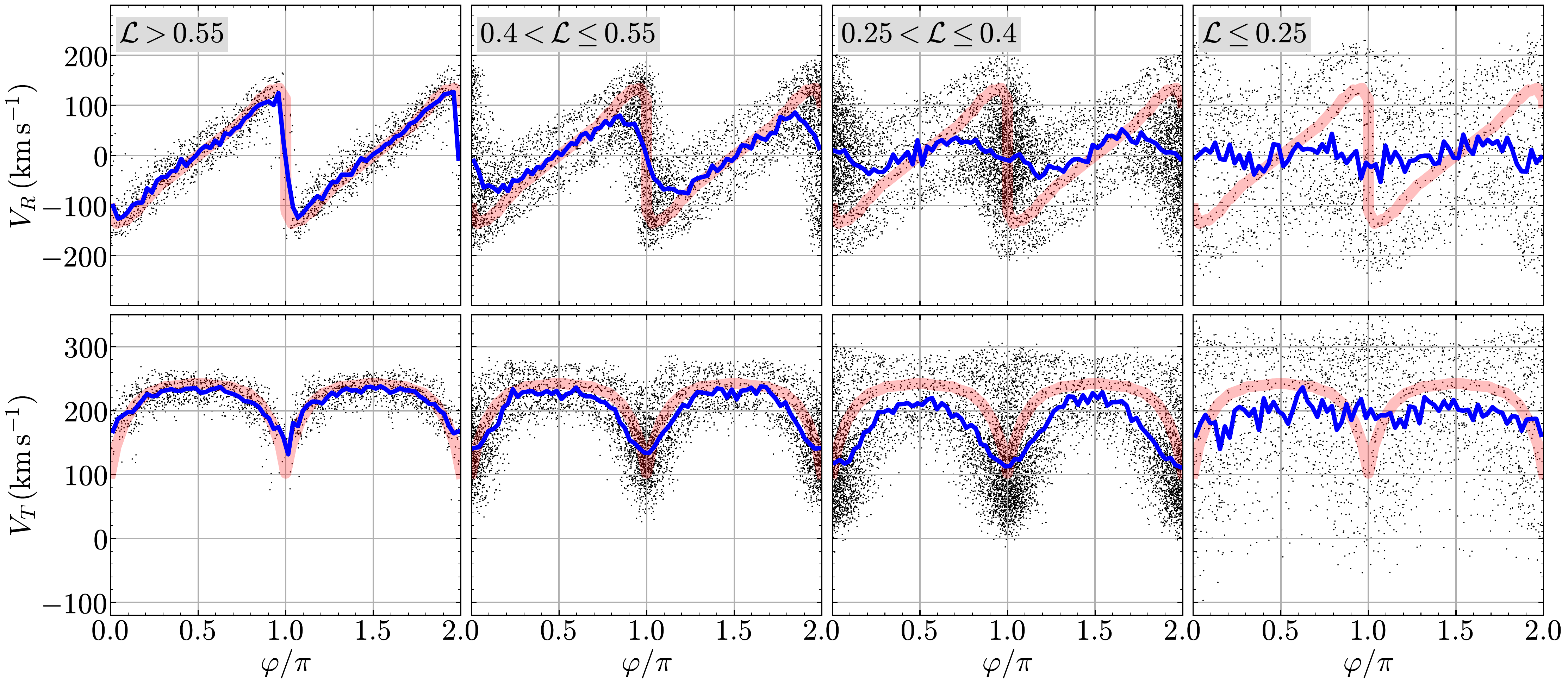}
\caption{Angular distributions of (upper) radial and (lower) tangential velocities of disk particles on bar-supporting orbits with $A_x/A_y\geq1.5$, plotted as dots, located at $R=0.5R_b$ in model {\tt C20} at $t=6 \Gyr$. The panels, from left to right, correspond to decreasing ranges of normalized mean angular momentum: $\mathcal{L}>0.55$, $0.55\geq \mathcal{L}>0.40$, $0.40\geq \mathcal{L}>0.25$, and $\mathcal{L}\leq 0.25$.  Blue lines indicate the boxcar-averaged velocities with a window size of $\Delta \phi = 5^{\circ}$, while thick red curves represent the characteristic sawtooth-like and arch-like patterns of the $x_1$ orbits, respectively.}
\label{fig:bar_supporting_orbits}
\end{figure*}

\begin{figure*}[t]
\plotone{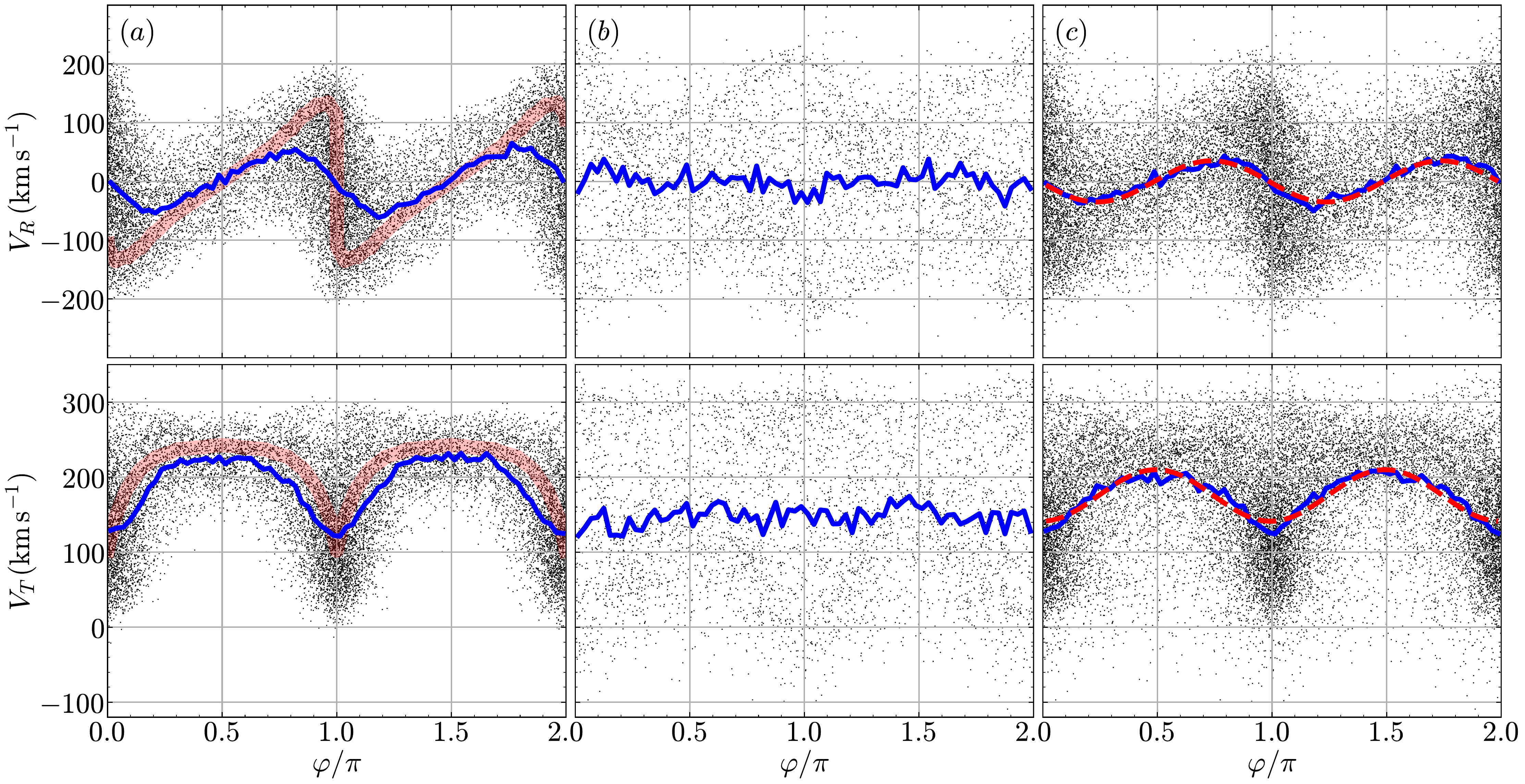}
\caption{Angular distributions of (upper) radial and (lower) tangential velocities for disk particles located at $R=0.5R_b$ in model {\tt C20} at $t=6\Gyr$. Dots in each panel represent the velocities of particles on ($a$) $x_1$-like orbits, ($b$) box and disk orbits, and ($c$) all orbits. Blue lines indicate their boxcar-averaged velocities, computed with a window size of $\Delta \phi = 5^\circ$. In $(a)$, the thick red curves depict the characteristic sawtooth-like and arch-like patterns of the $x_1$ orbits. In ($c$), the red dashed lines plot \cref{eq:bisymmetric0} using appropriately chosen coefficients.
}
\label{fig:bar_and_disk}
\end{figure*}

\subsection{$N$-body Bars}\label{sec:bisymmetric_Nbody}

Given that sawtooth-like and arch-like patterns are inherent to the velocity distributions of $x_1$ orbits, it is interesting to investigate whether bars in $N$-body simulations also exhibit similar velocity patterns.
\cref{fig:bisymmetric_C10,fig:bisymmetric_C20} plot the face-on views of the disk surface density $\Sigma$ and the angular distributions of the particle velocities for models {\tt C10} and {\tt C20} at $t=4$ and $6\Gyr$, respectively. The dots in the middle panels indicate the radial and tangential velocities of the particles in an annulus at $R=0.5R_{b}$ with thickness $\Delta R=0.1\kpc$, while the thick red lines correspond to the velocities of particles on the $x_1$ orbits, as shown in \cref{fig:frozen_x1_velocity}. Despite variations in bar strength and morphology, a significant fraction of particles follow the characteristic sawtooth-like and arch-like velocity patterns of $x_1$ orbits in all snapshots containing a bar, although some exhibit deviations from typical $x_1$ kinematics.

The right panels of \cref{fig:bisymmetric_C10,fig:bisymmetric_C20} plot the boxcar-averaged mean values and standard deviations, computed with a window width of $5^\circ$, as black lines and shaded regions, respectively. Also plotted as red dashed curves is the bisymmetric model, \cref{eq:bisymmetric0}, with coefficients derived from the Fourier transforms of the mean profiles. The close resemblance between the black solid and red dashed lines indicates that the bisymmetric model provides an accurate description of the angular distributions of stellar velocities, but only in an averaged sense. It captures neither the characteristic sawtooth-like and arch-like patterns of $x_1$ orbits nor the large dispersions in velocity distributions, amounting to $\sim 50$--$100 \kms$.

\begin{figure}[t]
\plotone{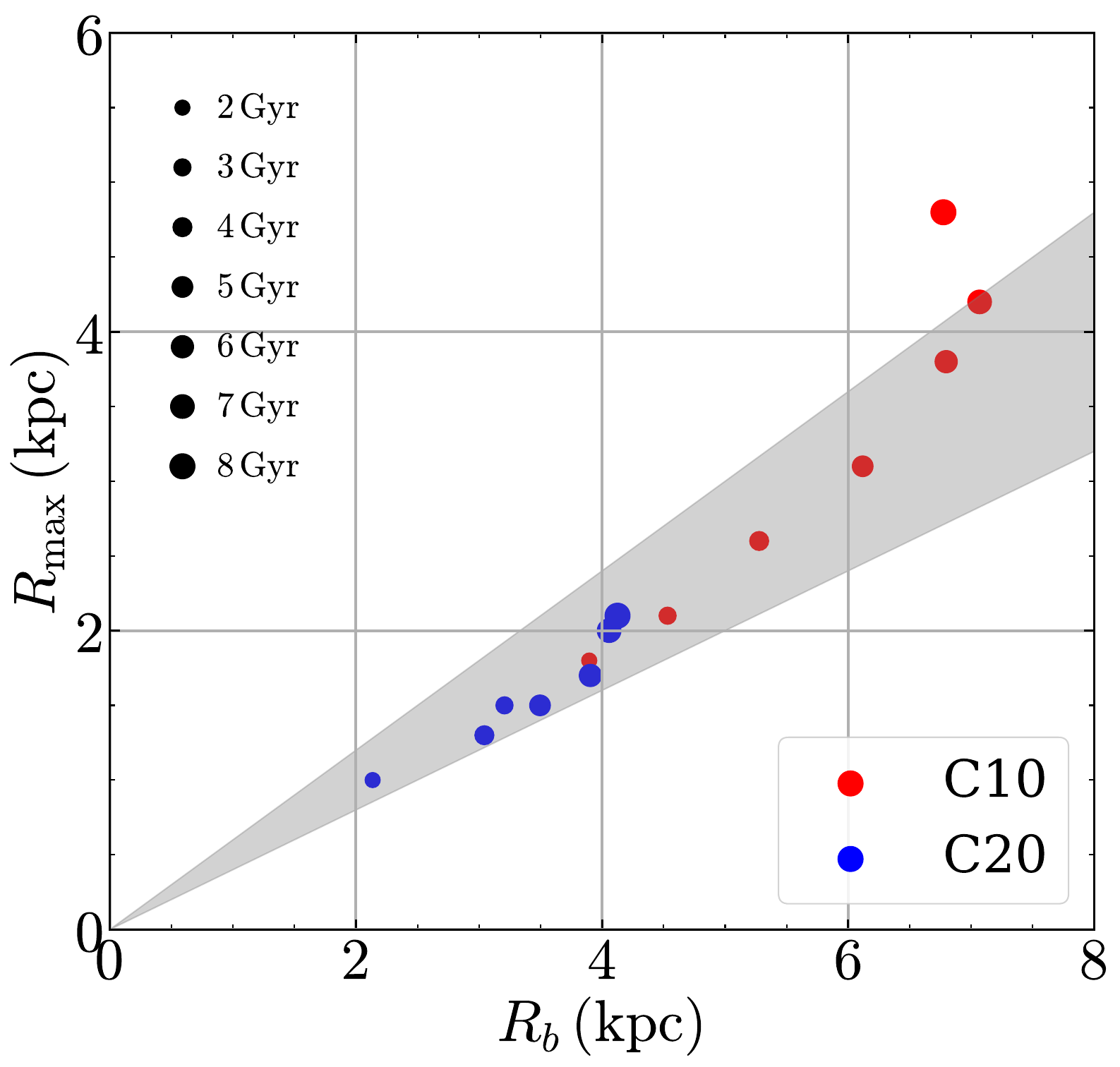}
\caption{Relation between the bar length $R_b$ and the radius $R_\text{max}$ at which the bisymmetric noncircular motion amplitude $A_\text{bis}(R)$ reaches its maximum. The shaded regions represent  $R_\text{max} = (0.4$--$0.6)R_b$.}
\label{fig:Rmax}
\end{figure}

To better understand the origin of the similarity between the averaged velocity distribution and the bisymmetric model, we examine the velocity distributions of individual orbital groups.
\Cref{fig:bar_supporting_orbits} plots the $\mathcal{L}$-dependence of the angular distributions of radial and tangential velocities for disk particles on bar-supporting orbits with $A_x/A_y \ge 1.5$ at a radius of $R=0.5R_b$ in model {\tt C20} at $t=6\Gyr$. The panels, from left to right, correspond to $\mathcal{L}>0.55$, $0.55\geq \mathcal{L}>0.40$, $0.40\geq \mathcal{L}>0.25$, and $\mathcal{L}\leq 0.25$. 
Particles on bar-supporting orbits with $\mathcal{L} > 0.55$ closely follow the characteristic sawtooth-like radial and arch-shaped tangential velocity patterns of ideal $x_1$ orbits. In contrast, those on $x_1$-like orbits with lower $\mathcal{L}$ progressively deviate from these idealized trajectories, leading to a blurring of the sawtooth and arch-shaped features in the averaged velocity profiles. This deviation becomes increasingly pronounced as $\mathcal{L}$ decreases. Particles on box orbits with $\mathcal{L} \leq 0.25$ as well as particles on disk orbits show no trace of these characteristic patterns.

\begin{figure}[t]
\plotone{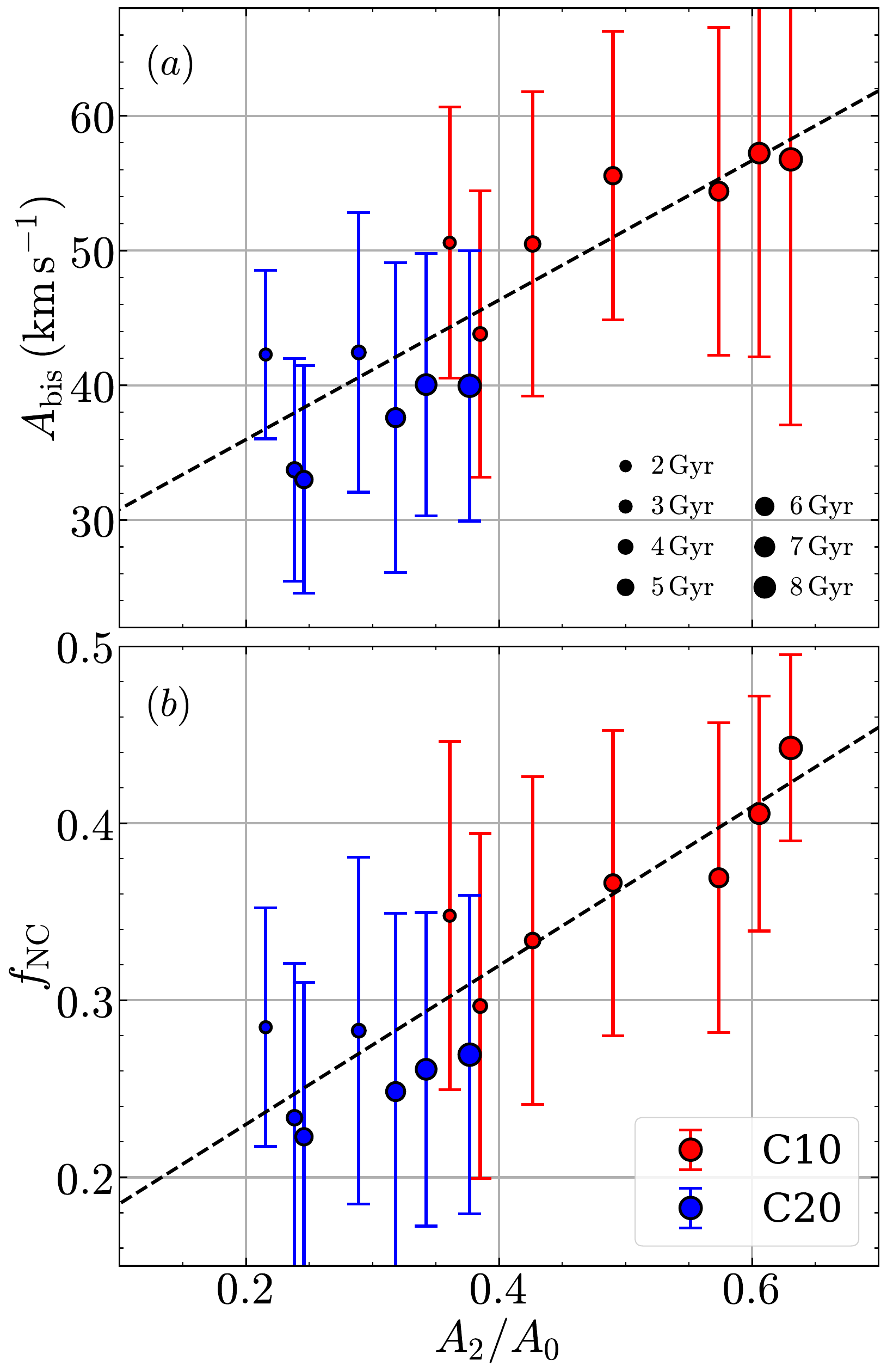}
\caption{Mean values of ($a$) the bisymmetric noncircular motion amplitude $A_{\mathrm{bis}}$ and ($b$) its fractional contribution relative to the circular rotation velocity, $f_{\mathrm{NC}}$, averaged over the radial range $0.2 \le R/R_b \le 0.8$, as functions of the bar strength $A_{2}/A_{0}$ for models {\tt C10} (red) and {\tt C20} (blue). Error bars represent standard deviations. Marker sizes indicate the simulation times. The black dashed line indicates the best-fit relations, \cref{eq:Abisfit,eq:fNCfit}.}
\label{fig:A_bis_correlation}
\end{figure}

\cref{fig:bar_and_disk} plots the angular distributions of radial and tangential velocities for disk particles on $x_1$-like orbits, box and disk orbits, and all orbits, separately, at $R=0.5R_b$ in model {\tt C20} at $t=6\Gyr$. Thin solid lines indicate the boxcar-averaged velocities, computed with a window size of $\Delta \phi=5^\circ$.  Clearly, particles on $x_1$-like orbits retain the characteristic sawtooth and arch-shaped patterns, albeit with slightly reduced amplitudes due to the presence of particles with low $\mathcal{L}$. In contrast, particles on box or disk orbits have velocities that show no significant azimuthal variation, on average. Their motions can be considered effectively random, further reducing the prominence of the sawtooth-like and arch-shaped patterns in the azimuthal distributions of the averaged velocities of all particles.

As shown in the right panels of \cref{fig:bar_and_disk}, the averaged profiles of radial and tangential velocities for all particles closely match the predictions of the bisymmetric model. This agreement emerges because the characteristic sawtooth-like and arch-shaped patterns, prominent in $x_1$-like orbits with high $\mathcal{L}$, are gradually suppressed as particles on $x_1$-like orbits with lower $\mathcal{L}$ and those on box and disk orbits are included. As the contribution from low-$\mathcal{L}$ particles increases, the sharp features associated with $x_1$ orbits become less pronounced, resulting in smoother velocity profiles that are well described by a sinusoidal variation with $m = 2$ symmetry.

\begin{figure}[t!]
\plotone{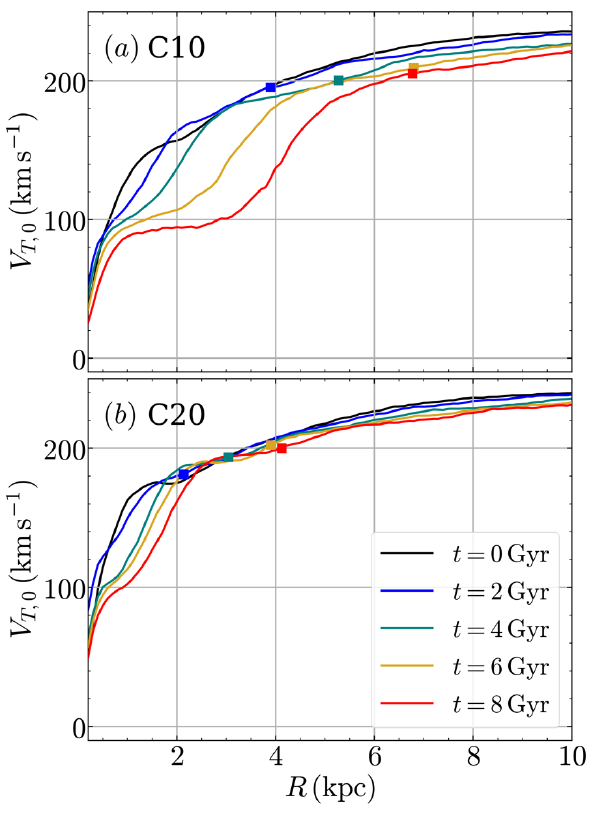}
\caption{Temporal evolution of the rotation curves $V_{T,0}(R)$ in ($a$) model {\tt C10} and ($b$) model {\tt C20}. Squares indicate the bar length, $R = R_b$, at times $t \geq 2 \Gyr$.}
\label{fig:rotation_curve}
\end{figure}

\subsection{Bisymmetric Noncircular Motions}

Given that the bisymmetric model captures the azimuthal distributions of the averaged stellar velocities, it is reasonable to define the amplitude of the bisymmetric perturbed motions as
\begin{equation}\label{eq:Abis}
A_\mathrm{bis}(R) = \sqrt{V_{R,2}^{2}(R) + V_{T,2}^{2}(R)}
\end{equation}
and to express its strength relative to the unperturbed rotational velocity via
\begin{equation}\label{eq:fNC}
f_\mathrm{NC}(R) = A_{\mathrm{bis}}(R)/V_{T,0}(R)
\end{equation}
\citep[see][]{LopezCoba22, Kim24}. 
\Cref{fig:Rmax} plots the radius $R_\text{max}$, at which $A_\mathrm{bis}$ attains its maximum, as a function of the bar length $R_b$, illustrating that the bars in our models roughly exhibit $R_\text{max} \sim 0.5 R_b$. This is broadly consistent with the result of \citet{Kim24}, who found that the amplitudes of noncircular motions in barred galaxies peak, on average, at $R \sim 0.45R_b$.

\begin{figure}[t]
\plotone{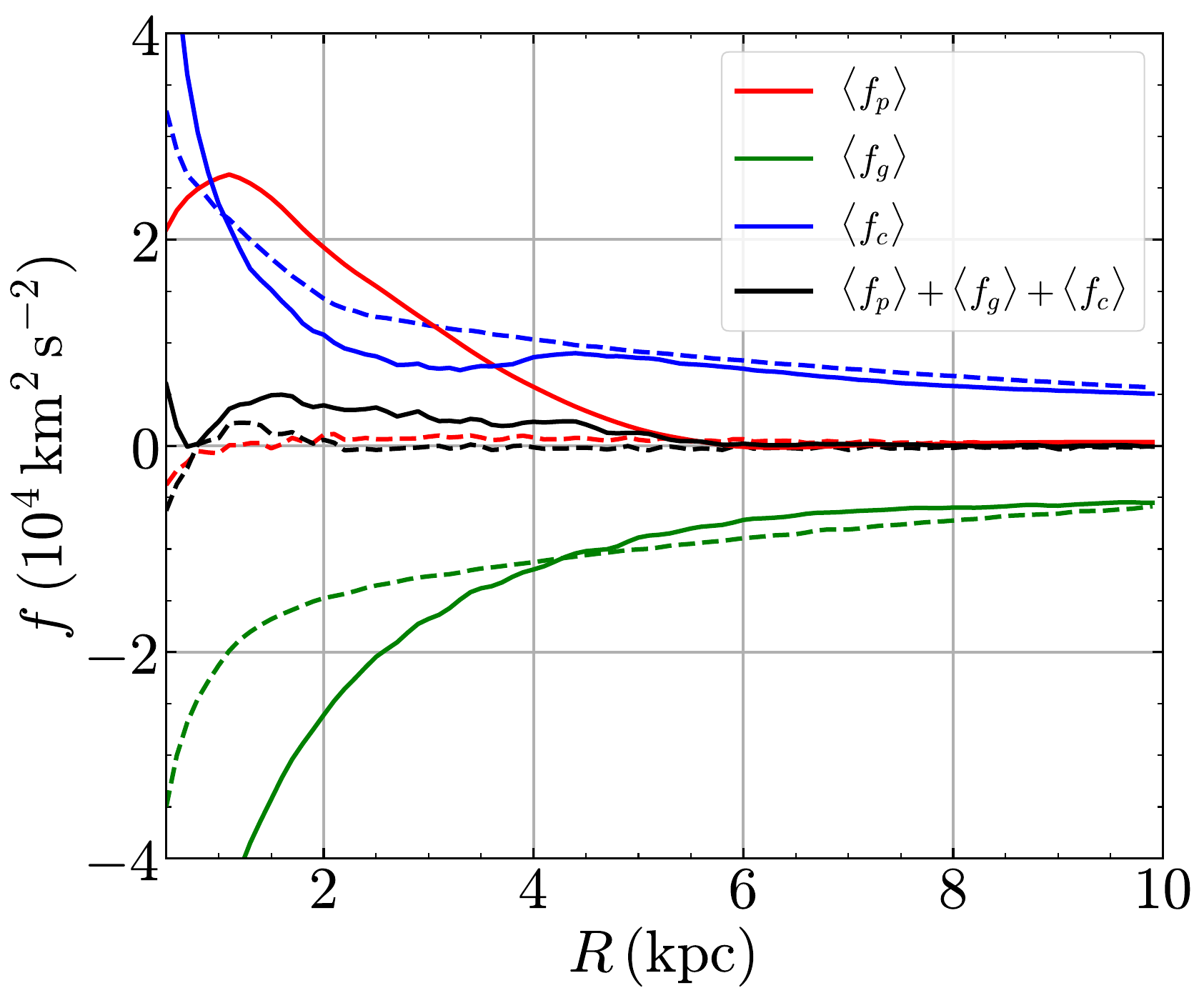}
\caption{Radial distributions of the azimuthally-averaged pressure gradient force (red), gravitational force (green), centrifugal force (blue), and net force (black) for model {\tt C10}. The dashed and solid lines correspond to the forces at $t = 0$ and $t = 8\Gyr$, respectively.}
\label{fig:Jeans}
\end{figure}

\cref{fig:A_bis_correlation} plots the mean values and standard deviations of $A_{\mathrm{bis}}$ and $f_{\mathrm{NC}}$, averaged over the radial range $0.2 \le R/R_b \le 0.8$, as functions of the bar strength $A_{2}/A_{0}$ for models {\tt C10} (red) and {\tt C20} (blue). It is evident that the velocity perturbations, as well as their fractional amplitudes relative to the unperturbed rotation, are strongly correlated with the bar strength. Our best-fit relations are
\begin{align}
 A_\text{bis} &= 52 (A_2/A_0) + 26 \quad\kms, \label{eq:Abisfit} \\
 f_{\text{NC}} &= 0.45 (A_2/A_0) + 0.14, \label{eq:fNCfit}
\end{align}
which are plotted as the dashed lines. Note that $A_\text{bis}$ is smaller by a factor of $\sim4$–$5$ compared to $\Delta V_R$ and $\Delta V_T$ in \cref{eq:x1fits}. This difference arises from the presence of stellar particles on box or disk orbits, which introduce random components to the perturbed velocities, thereby smearing out the sawtooth-like and arch-shaped patterns of $x_1$ orbits. Nevertheless, the strong correlations of $A_\text{bis}$ and $f_\text{NC}$ with $A_2/A_0$ support observational findings that bars are responsible for driving noncircular stellar motions in galactic disks \citep[e.g.,][]{LopezCoba22, Kim24}.

\begin{figure}[t]
\plotone{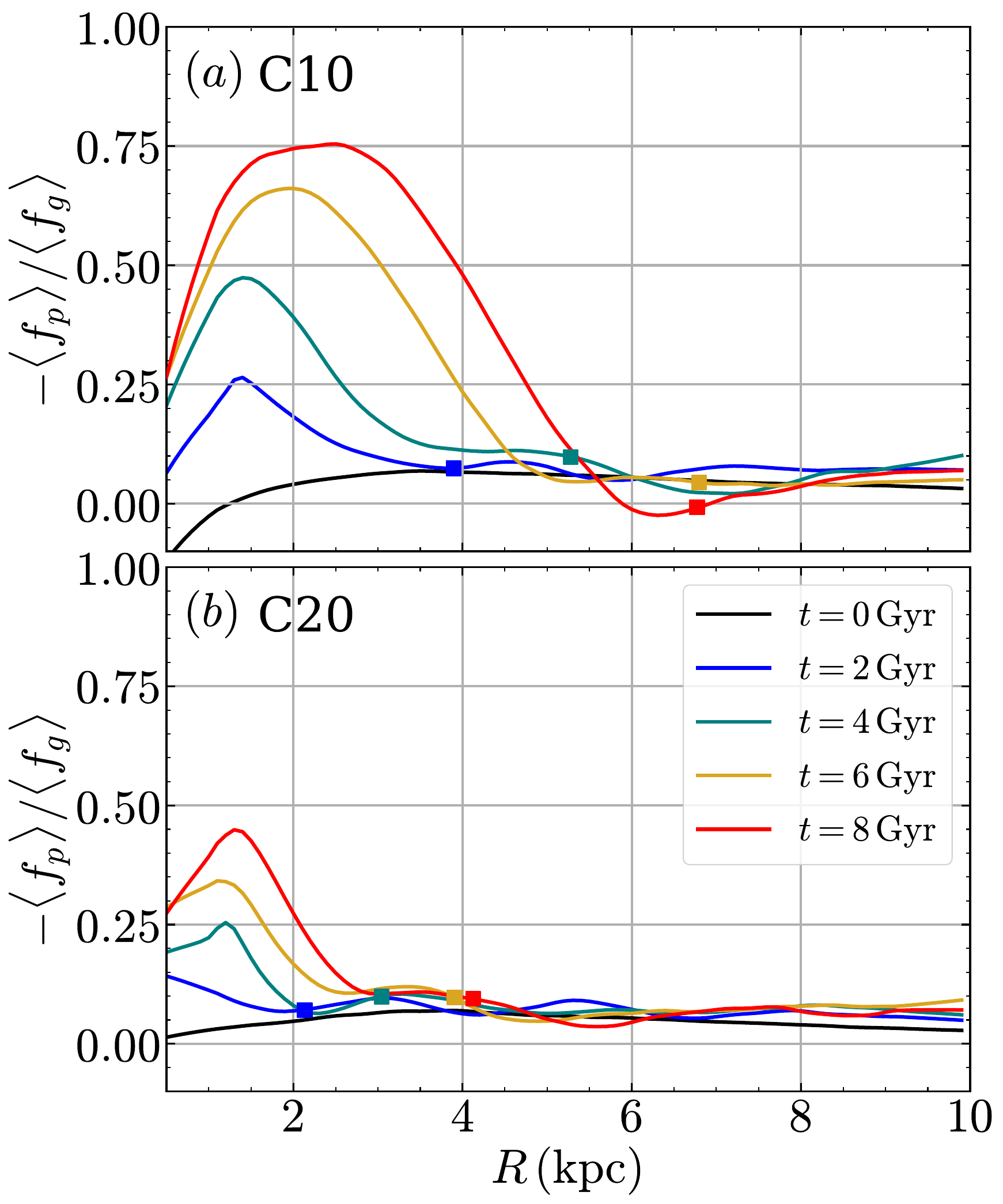}
\caption{Ratio of the azimuthally averaged pressure gradient force to gravitational force as functions of $R$ and $t$ for $(a)$ model {\tt C10} and $(b)$ model {\tt C20}. Squares indicate the bar length, $R = R_b$, at times $t \geq 2 \Gyr$.
}
\label{fig:ForceRatio}
\end{figure}

\section{Rotation Curve}\label{sec:rotation}

Since stellar velocities in barred galaxies are non-axisymmetric, it is reasonable to take azimuthal averages of the velocity field to derive rotation curves \citep[e.g.,][]{Athanassoula13}. In fact, $V_{T,0}(R)$ in \cref{eq:bisymmetric0}, which represents the $m=0$ Fourier component of the tangential velocity, corresponds to the rotational velocity at $R$.  \cref{fig:rotation_curve} plots the temporal variations of $V_{T,0}(R)$ in models {\tt C10} and {\tt C20}. In both models, the rotational velocity decreases in the bar region as the bar becomes longer and stronger. The reduction of $V_{T,0}(R)$ in model {\tt C10} is more pronounced and extends over a wider radial range compared to model {\tt C20}, reflecting the stronger and more extended bar structure in the former. This finding is consistent with the results of \citet{Bureau05}, who demonstrated that galaxies with stronger bars exhibit more prominent double-hump structures in their mean stellar velocity profiles along the \ac{LOS}.

What causes the reduction in rotational velocities in barred galaxies? Is it primarily a consequence of modified gravitational forces or enhanced stellar random motions? In an axisymmetric stellar disk, the radial force balance can be described by the Jeans equation \citep{BT08}:
\begin{align}\label{eq:Jeans}
 -\frac{1}{\nu} \frac{\partial (R \nu \overline{v_R^2})}{\partial R} - R \frac{\partial \Phi}{\partial R} + \overline{v_T^2} =0,
\end{align}
where $\nu$ is the number density of stars and $\Phi$ is the gravitational potential. Each term in \cref{eq:Jeans} corresponds, from left to right, to the pressure gradient force $f_p \equiv - (R\nu)^{-1} \partial (R \nu \overline{v_R^2})/\partial R$, the gravitational force $f_g \equiv - \partial \Phi / \partial R$, and the centrifugal force $f_c = \overline{v_T^2}/R$, all defined per unit mass at a cylindrical radius $R$.
The degree to which a system is rotation- or pressure-supported is determined by the ratio $|f_p/f_g|$.

\begin{figure}[t]
\plotone{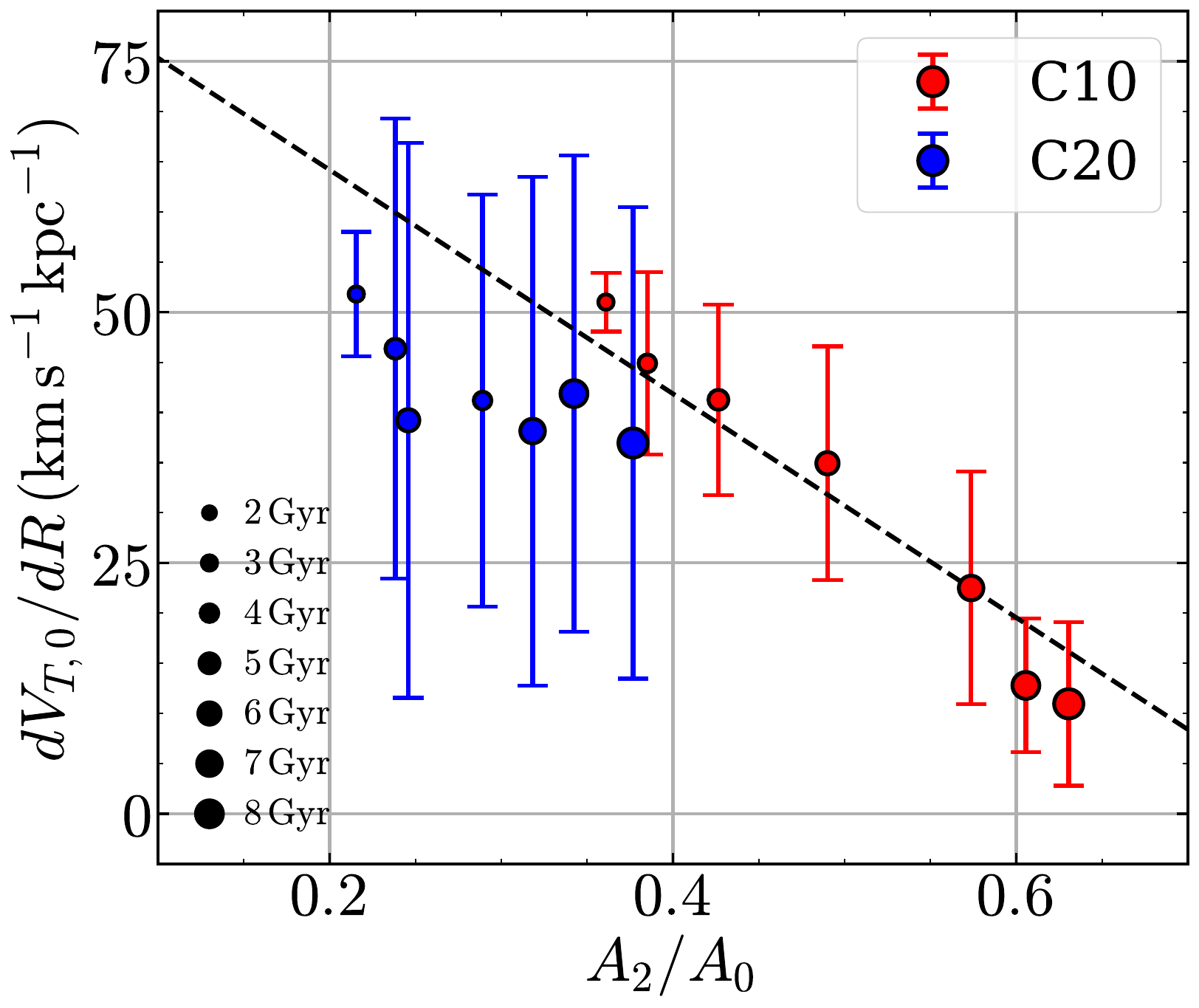}
\caption{Mean slopes of the rotation curves, averaged over the radial range $0.2 \leq R/R_b \leq 0.8$, as a function of bar strength $A_2/A_0$ for models {\tt C10} (red) and {\tt C20} (blue). Error bars represent standard deviations. The dashed line represents our best-fit relation, given by \cref{eq:slopefit}.}
\label{fig:slope}
\end{figure}

Since a barred galaxy is not axisymmetric, we take the azimuthal average of each term in \cref{eq:Jeans} and examine the extent to which the equation holds for the averaged quantities. \cref{fig:Jeans} plots the spatial profiles of the azimuthally-averaged terms for model {\tt C10} at $t=0$ (dashed) and $8\Gyr$ (solid).
\cref{fig:ForceRatio} plots the radial profiles of $-\langle f_p \rangle/\langle f_g\rangle$ for models {\tt C10} and {\tt C20} at various times, where $\langle\,\rangle$ indicates averaging over azimuth. At $t=0\Gyr$, the radial pressure gradient is negligible across the disk, suggesting that the system is initially dominated by rotational support. As the bar forms and grows, the radial component of the stellar velocities increases, enhancing the pressure gradient force. Simultaneously, the non-axisymmetric gravitational forces induced by the bar interact resonantly with stars near the corotation and Lindblad resonances \citep{SW93, BT08}. This interaction drives inward stellar migration \citep[e.g.,][]{Minchev10}, thereby increasing the gravitational force. In the bar region, the increase in $\langle f_p \rangle$ exceeds that in $\langle -f_g \rangle$, resulting in a reduction of $\langle f_c \rangle$ to maintain equilibrium. 
In other words, the reduced rotational velocity in the bar region, as shown in \cref{fig:rotation_curve}, results from enhanced random stellar motions induced by the bar. \Cref{fig:ForceRatio} shows that $\langle f_p \rangle$ increases with bar strength and that regions of high $-\langle f_p \rangle/\langle f_g\rangle$ correspond to regions of low rotational velocity displayed in \cref{fig:rotation_curve}. The contribution of pressure support to the force balance becomes significant in the presence of a strong bar.

\cref{fig:rotation_curve} suggests that the rotation curve in the bar region becomes shallower with increasing bar strength. To quantify the slope, we apply a boxcar smoothing to the rotation curves using a window size of $\Delta R = 0.5\kpc$, and measure the mean gradient $dV_{T,0}/dR$ over the radial range $0.2 \leq R/R_b \leq 0.8$.
\cref{fig:slope} displays the relation between the mean velocity gradient $dV_{T,0}/dR$ and the bar strength $A_2/A_0$. Error bars indicate standard deviations. Galaxies with stronger bars clearly exhibit shallower rotation curves within the bar region. The best-fit linear relation is given by
\begin{equation}\label{eq:slopefit}
\frac{dV_{T,0}}{dR} = -112 \left(\frac{A_2}{A_0}\right) + 86 \;\kms \, \kpc^{-1},
\end{equation}
and is plotted as the dashed line. This trend suggests a potential indirect method for inferring bar strength from kinematic measurements.

\section{Discussion}\label{sec:discussion}

In this section, we discuss our findings in the context of decomposing the \ac{LOS} velocities of observed galaxies using {\tt XookSuut}.
We also discuss the impact of bars on stellar noncircular motions and rotation curves within the bar region in comparison with observed barred galaxies.

\subsection{Bisymmetric Model Fitting}

We have shown that while stellar particles on bar-supporting $x_1$ orbits exhibit characteristic sawtooth-like and arch-like patterns in the azimuthal distributions of their radial and tangential velocities, the presence of box and disk orbits produces averaged profiles that are well described by the bisymmetric model proposed by \citet{SS07}. In fact, the bisymmetric model has been widely adopted in observational studies of barred galaxies \citep[e.g.,][]{Hallenbeck14, Holmes15, LopezCoba22, Zanger24, Hogarth24, Kim24, LopezCoba24a, LopezCoba24b}. Given that the model is valid only in an averaged sense, the reliability of bar-induced velocity perturbations inferred from observations remains uncertain. To address this, we employ our simulation data to generate mock observations and evaluate the accuracy of the bisymmetric model in measuring the velocity perturbations.

\begin{figure}[t]
\plotone{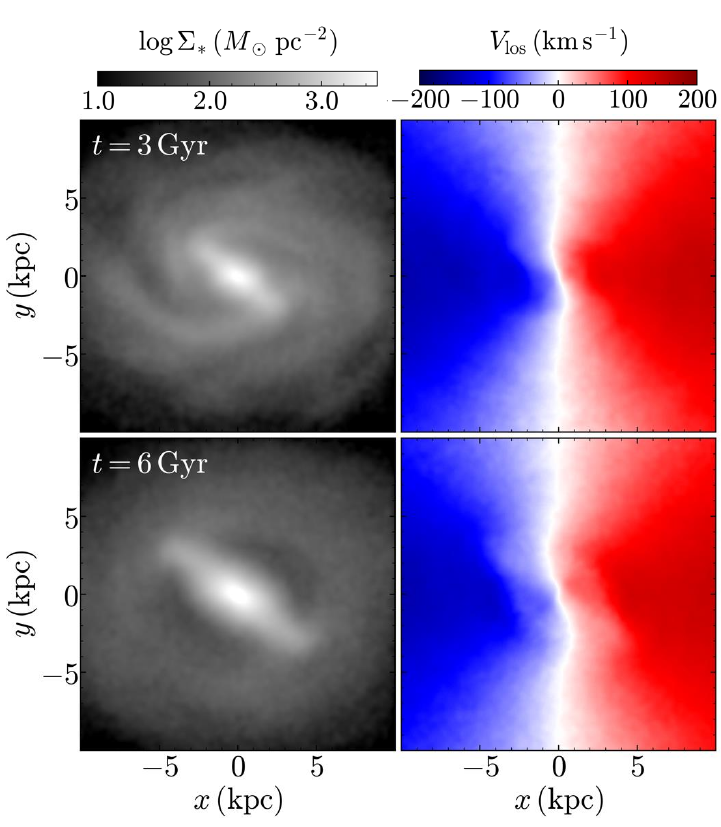}
\caption{Distributions on the sky of (left) the projected density $\Sigma_*$ and (right) the \ac{LOS} velocity $V_\text{los}$ for the model galaxies at (top) $t=3\Gyr$ and (bottom) $t=6\Gyr$. The disk is inclined by $i = 40^\circ$, and the bar \ac{PA} is set to $\phi_b = 135^\circ$. The line of nodes lies along the $y=0$ axis.}
\label{fig:Imgal}
\end{figure}

For this purpose, we select two datasets from model {\tt C10} at $t = 3$ and $6 \Gyr$, rotate them to align the \ac{PA} of the bar's semimajor axis with $\phi_b = 135^\circ$, and incline them by $i = 40^\circ$ with respect to the plane of the sky. We observe the resulting galaxies using a Gaussian beam $\Theta$ with a full width at half maximum of $0.4\kpc$, representative of the spatial resolution typical of MUSE datasets \citep{LopezCoba20}. We calculate the projected density  $\Sigma_* = \int \Theta \rho dxdydz/\int \Theta dxdy$ and the \ac{LOS} velocity $V_\text{los}=\int \Theta \rho v_\text{los} dz /\Sigma_*$, where $dxdy$ denotes the area element in the plane of the sky and $dz$ is the length element along the \ac{LOS}, and $v_{\text{los}}$ is the \ac{LOS} velocity, defined as 
\begin{equation}
v_{\text{los}} = \sin i( v_T \cos \theta + v_R \sin \theta) - v_z \cos i,
\end{equation}
where $v_R\equiv v_x\cos\theta + v_y \sin\theta$, $v_T\equiv -v_x \sin\theta + v_y \cos\theta$, and  
$\theta$ is the azimuthal angle measured from the line of nodes. \cref{fig:Imgal} plots the resulting distributions of $\Sigma_*$ and $V_\text{los}$ for the two model galaxies. The characteristic S-shaped feature is clearly visible near the center of the $V_\text{los}$ maps \citep[e.g.,][]{Peterson78, Bosma78, Huntley78, Peterson80, Pence81, Kormendy83, Pence84, Weliachew88, Fathi05, Stark18}, and becomes more prominent at $t = 6 \Gyr$, when the bar is stronger.

\begin{figure*}[t]
\plotone{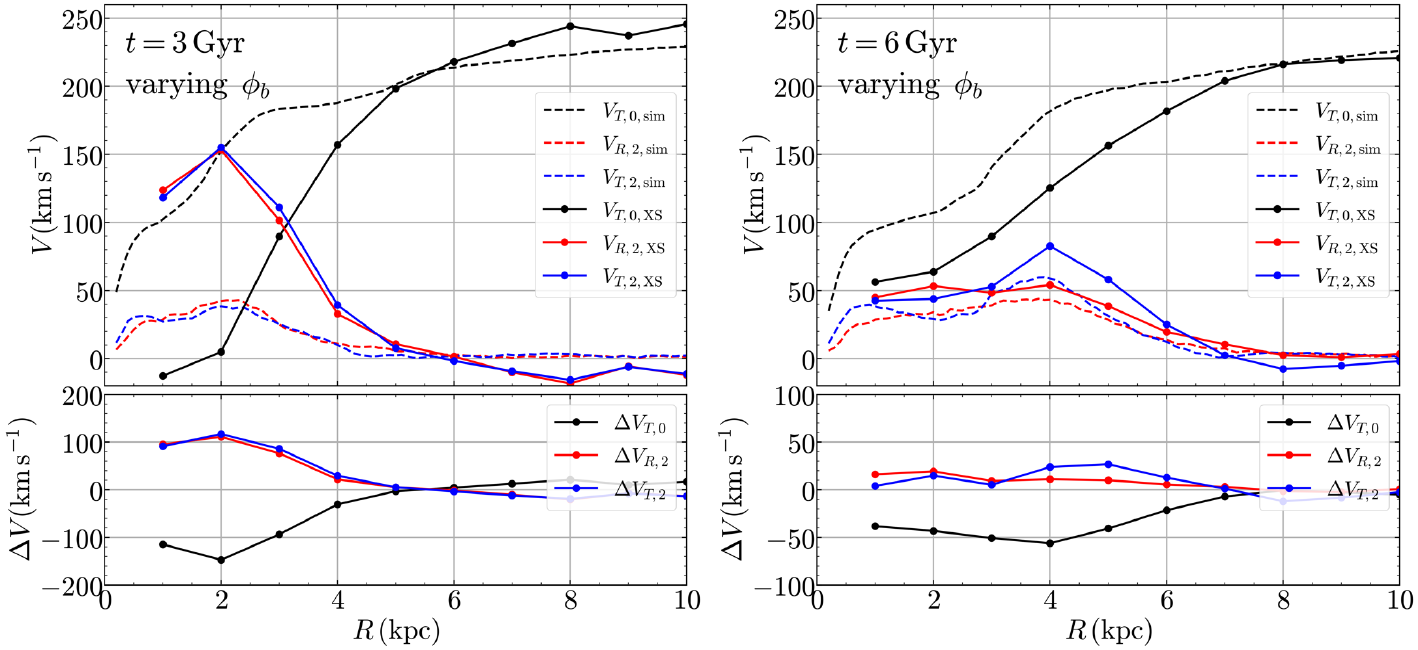}
\caption{Results of the velocity decomposition using {\tt XookSuut}, with the bar \ac{PA} determined internally, for the galaxies at (left) $t=3\Gyr$ and (right) at $t=6\Gyr$. The upper panels show the {\tt XookSuut} results, $V_\text{XS}$, as solid lines, compared with the original simulation data, $V_\text{sim}$, shown as dashed lines. The lower panels display the difference, $\Delta V=V_\text{XS}-V_\text{sim}$.
}
\label{fig:Velfit1}
\end{figure*}

\begin{figure*}[t!]
\plotone{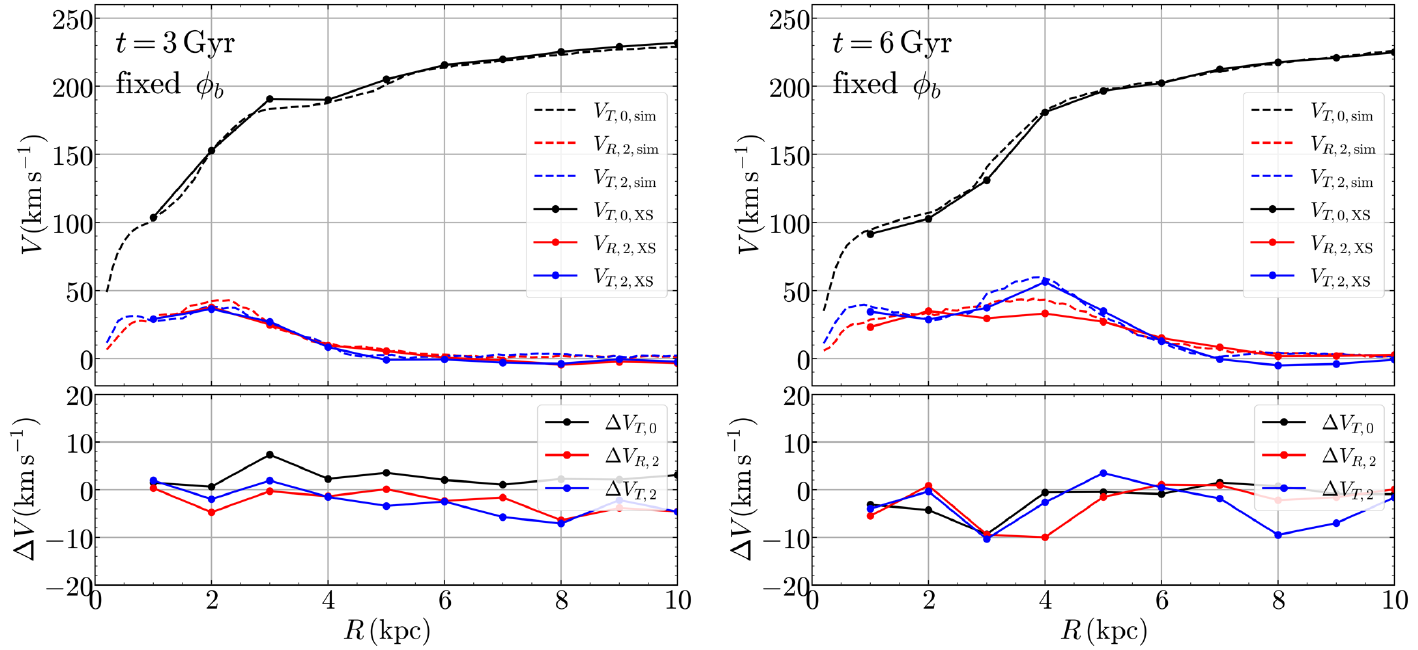}
\caption{Same as \cref{fig:Velfit1}, but with the bar \ac{PA} fixed at $\phi_b=135^\circ$.}
\label{fig:Velfit2}
\end{figure*}

We proceed to decompose $V_\text{los}$ into $V_{T,0}$, $V_{T,2}$, and $V_{R,2}$ within the framework of the bisymmetric model, utilizing the {\tt XookSuut} code \citep{LopezCoba24c}. The current version of {\tt XookSuut} is configured to determine the bar \ac{PA} in addition to the various velocity components as part of the least-squares and Markov Chain Monte Carlo fitting processes. \cref{fig:Velfit1} plots the resulting decomposed velocities $V_\text{XS}$, in comparison with $V_\text{sim}$ obtained from the Fourier analysis of the original simulation data, as well as their difference, $\Delta V=V_\text{XS}-V_\text{sim}$. The velocity decomposition obtained with {\tt XookSuut} appears unreliable, exhibiting significant deviations from the original simulation data in both model galaxies. At $t = 6\Gyr$, the model shows errors of approximately $\Delta V_{T,0} \sim 50\kms$ in the rotational velocity, and $\Delta V_{R,2} \sim 20\kms$ and $\Delta V_{T,2} \sim 30\kms$ in the perturbed components. The discrepancies are even more pronounced at $t = 3\Gyr$, reaching up to $|\Delta V| \sim 100\kms$. The source of these large deviations remains uncertain but is likely related to incorrect determination of the bar \ac{PA}, as the estimated values, $\phi_b = 97.1^\circ \pm 0.1^\circ$ and $109.9^\circ \pm 0.3^\circ$ at $t=3$ and $6\Gyr$, respectively, differ substantially from the true bar orientation.

\begin{deluxetable*}{lccccc}
\tablecaption{Properties of Sample Galaxies\label{tbl:gal}}
\tablehead{
\colhead{Name} & \colhead{$A_2/A_0$} & \colhead{$d$} & \colhead{$R_b$} & \colhead{$f_{\text{NC}}$} & \colhead{$dV_\text{rot}/dR$} \\ 
\colhead{} & \colhead{} & \colhead{(Mpc)} & \colhead{(kpc)} & \colhead{} & \colhead{(km s$^{-1}$ kpc$^{-1}$)} \\
\colhead{(1)} & \colhead{(2)} & \colhead{(3)} & \colhead{(4)} & \colhead{(5)} & \colhead{(6)}
}
\startdata
ESO 018-G018 & 0.18 & 75.2 & 1.9 & 0.05 $\pm$ 0.23 & 113.6 $\pm$ 0.0 \\
ESO 325-G043 & 0.23 & 179.7 & 5.7 & 0.04 $\pm$ 0.01 & 51.8 $\pm$ 11.8 \\
ESO 476-G016 & 0.17 & 79.3 & 6.1 & 0.34 $\pm$ 0.14 & 27.6 $\pm$ 13.7 \\
IC 2151      & 0.29 & 29.3 & 1.0 & 0.68 $\pm$ 0.20 & 21.8 $\pm$ 4.4 \\
IC 2160      & 0.31 & 55.3 & 5.3 & 0.23 $\pm$ 0.07 & 28.2 $\pm$ 30.4 \\
NGC 0289     & 0.16 & 19.2 & 1.9 & 0.60 $\pm$ 0.04 & 206.9 $\pm$ 125.6 \\
NGC 0692     & 0.20 & 91.4 & 5.0 & 0.25 $\pm$ 0.03 & 46.0 $\pm$ 27.9 \\
NGC 1591     & 0.20 & 48.2 & 2.4 & 0.29 $\pm$ 0.39 & 57.3 $\pm$ 3.9 \\
NGC 3464     & 0.11 & 44.6 & 4.8 & 0.09 $\pm$ 0.07 & 31.6 $\pm$ 17.0 \\
NGC 5339     & 0.24 & 44.6 & 5.6 & 0.20 $\pm$ 0.03 & 43.4 $\pm$ 20.2 \\
NGC 6947     & 0.23 & 71.2 & 9.9 & 0.14 $\pm$ 0.09 & 18.9 $\pm$ 22.0 \\
NGC 7780     & 0.21 & 74.0 & 6.5 & 0.19 $\pm$ 0.10 & 29.5 $\pm$ 11.0 \\
PGC 055442   & 0.11 & 105.5 & 2.8 & 0.19 $\pm$ 0.02 & 73.6 $\pm$ 71.2 \\
UGC 03634    & 0.19 & 108.0 & 6.7 & 0.13 $\pm$ 0.03 & 51.3 $\pm$ 22.9
\enddata
\tablecomments{Column (1): galaxy name; Column (2): bar strength; Column (3): distance to galaxy; Column (4): bar length; Column (5): ratio of the bisymmetric noncircular velocity to the rotational velocity; Column (6): slope of the rotation curve}
\end{deluxetable*}

Noting the bar \ac{PA} can be reliably estimated from a projected density (or photometric) map on the sky, we reapply {\tt XookSuut} to the model galaxies, this time fixing $\phi_b=135^\circ$. \cref{fig:Velfit2} plots the resulting velocity decomposition in comparison with the original data. It is worth noting that the velocity errors are $|\Delta V| \lesssim 10\kms$ for both model galaxies, which are substantially smaller than those obtained when $\phi_b$ is determined internally. This suggests that using the bar \ac{PA} inferred from photometric data, rather than treating it as a free parameter, yields a more reliable velocity decomposition.

We emphasize that the perturbed velocities obtained using {\tt XookSuut} represent only the mean velocities. Since they lack the characteristic sawtooth-like and arch-like patterns (see \autoref{fig:bar_and_disk}), they do not reliably represent the velocity perturbations of stars on bar-supporting $x_1$ orbits. As shown in \cref{fig:bisymmetric_C10,fig:bisymmetric_C20}, the associated velocity dispersions are on the order of $\sim50$--$100\kms$, comparable in magnitude to the perturbed mean velocities. Since the bisymmetric model is, by construction, unable to capture velocity dispersions, it is desirable to develop a model that can simultaneously recover both the mean velocities and the velocity dispersions.

\subsection{Bisymmetric Noncircular Motions}

\begin{figure}[t]
\plotone{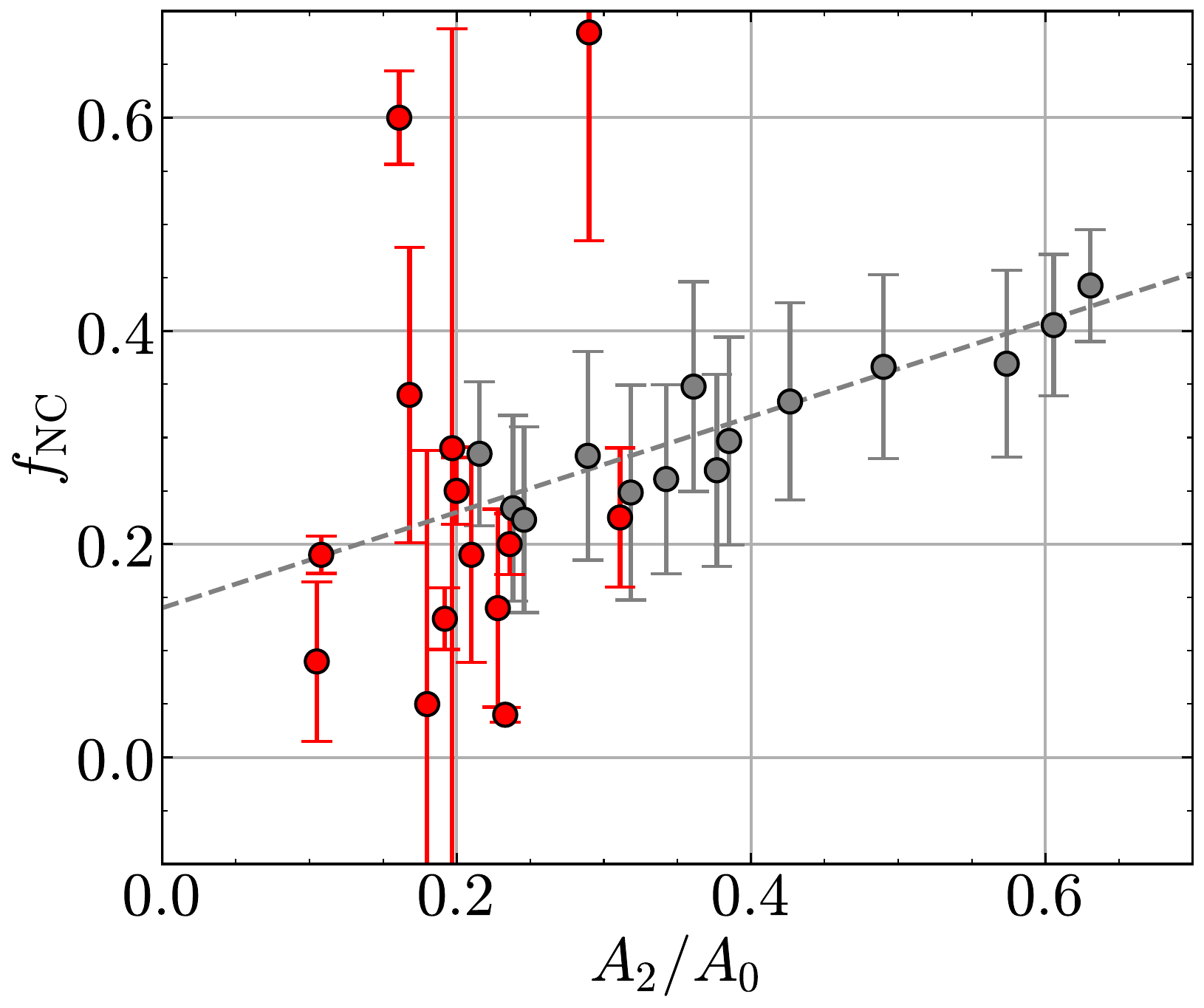}
\caption{Ratios of the bisymmetric noncircular motions to the circular rotation, $f_{\mathrm{NC}}$, as a function of bar strength, $A_{2}/A_{0}$. Red circles represent a sample of 14 observed galaxies analyzed by \citet{LopezCoba22} and \citet{Kim24}, while gray circles and dashed line correspond to the simulated galaxies, reproduced from \cref{fig:A_bis_correlation}.
}
\label{fig:fNC_obs}
\end{figure}

We have shown that the amplitude of bisymmetric noncircular motions, $A_{\text{bis}}$, and their fractional strength relative to circular rotation, $f_{\text{NC}}$, are strongly correlated with the bar strength when averaged within the bar region. To facilitate comparison with observations, we use the bisymmetric model fitting results of \citet{LopezCoba22}, obtained with {\tt XookSuut}, along with bar parameters from \citet{Kim24} for a sample of 14 barred galaxies. \cref{tbl:gal} lists the properties of the sample galaxies, along with the mean and standard deviation of $f_\text{NC}$ in Column (5), averaged over $0.2\leq R/R_b\leq 0.8$. To convert angular scales to physical units, we adopt galaxy distances from the NASA/IPAC Extragalactic Database\footnote{\url{https://ned.ipac.caltech.edu/}}. \cref{fig:fNC_obs} plots the relationship between $f_{\text{NC}}$ and $A_2/A_0$ for the observed galaxies, compared to the simulation results. Overall, the observed galaxies tend to have weaker bars than those in the simulations. Excluding two outliers (IC 2151 and NGC 0289), the observed trend of increasing $f_{\mathrm{NC}}$ with $A_2/A_0$ is broadly consistent with the simulation results.
The observed scatter seen here and in \cref{fig:slope_obs} may arise from the influence of other structural components (e.g., spiral arms) and uncertainties in the measurement of bar strength \citep{Kim24}. Given the limited sample size, further studies with statistically more complete datasets are needed to confirm and refine these findings.

It is also worth noting that two galaxies in our sample deviate noticeably from the general trend and may represent special cases. For instance, NGC 0289 has a bar \ac{PA} closely aligned with the disk \ac{PA}, which may introduce significant uncertainties in the fitting procedure \citep{SS07}. In addition, IC 2151 exhibits a relatively low $V_{T,0}$ within the bar region, likely resulting in an overestimated value of $f_{\mathrm{NC}}$.

\subsection{Rotation Curves}

We have also seen that the rotation curve $V_{T,0}(R)$ in the bar region decreases as the bar grows in size and strength. This reduction in $V_{T,0}(R)$ is attributed to increased random motions (or pressure gradient) in the radial direction. One may naively interpret the reduction in $V_{T,0}(R)$ as a rough indicator of bar strength. However, a few limitations make this approach challenging in observational contexts. First, observations of barred galaxies do not provide access to the rotation curve of the initial axisymmetric state, making it difficult to quantify the depth and extent of the velocity reduction. Second, observed photometric and spectroscopic data may include contributions from a classical bulge, if present, which is pressure-supported and tends to lower $V_{T,0}(R)$ in the central regions relative to the bulgeless case \citep[e.g.,][]{Beckman04}.

\begin{figure}[t]
\plotone{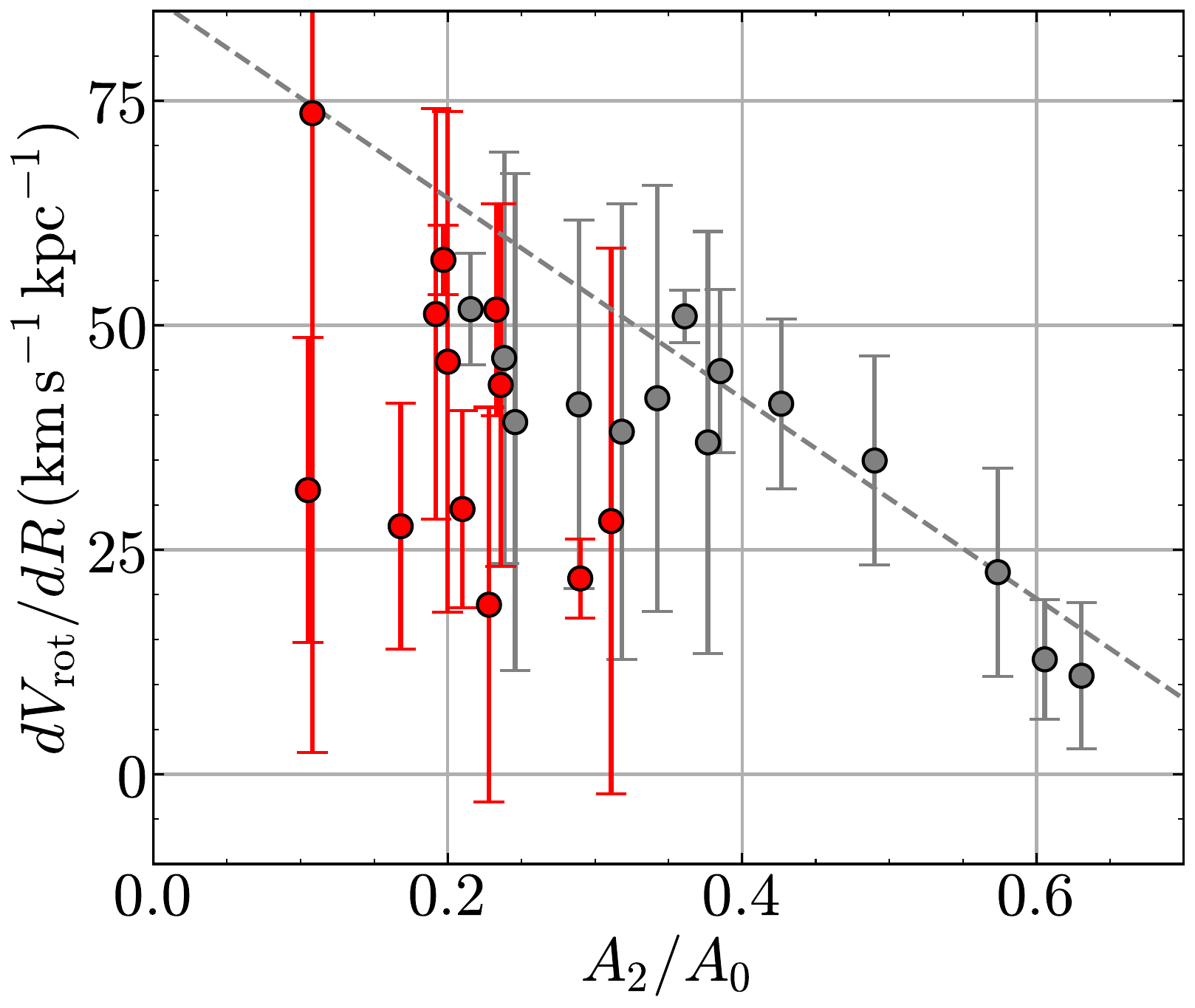}
\caption{Mean slopes of the rotation curves and their standard deviations, averaged over the radial range $0.2 \leq R/R_b \leq 0.8$, as a function of bar strength $A_2/A_0$. Red circles represent a sample of 12 galaxies analyzed by \citet{LopezCoba22} and \citet{Kim24}, while gray circles and dashed line are for the simulated galaxies, reproduced from \cref{fig:slope}.}
\label{fig:slope_obs}
\end{figure}

Nonetheless, \cref{fig:slope} suggests that, on average, galaxies with stronger bars tend to exhibit shallower rotation curves within the bar region. To examine whether this trend is also present in observations, we again use the results of \citet{LopezCoba22} and \citet{Kim24} to calculate the mean slopes $dV_\text{rot}/dR$ of the rotation curves for individual galaxies by averaging over the radial range $0.2 \leq R/R_b\leq 0.8$. The derived mean slopes and their standard deviations are presented in Column (6) of \cref{tbl:gal}. \cref{fig:slope_obs} plots the relation between $dV_\text{rot}/dR$ and $A_2/A_0$ for the observed galaxies, compared with the corresponding values from the simulated galaxies.\footnote{For clarity of presentation, \cref{fig:slope_obs} excludes the data points for galaxies ESO 018-G018 and NGC 0289, as they lie outside the range shown.} The slope of their rotation curves in the bar region shows a decreasing trend with bar strength, albeit weak, falling within a range not significantly different from that in the simulations.

\section{Summary}\label{sec:conclusion}

We have investigated two key aspects of the kinematic properties of barred galaxies: (1) the angular distributions of the radial and azimuthal components of stellar velocities, and (2) changes in the rotation curve induced by the presence of a bar. We select two Milky Way-sized galaxy models from \citet{JK23}, each containing a classical bulge with a mass equal to 10\% or 20\% of the disk mass. We analyze their $N$-body simulation data within the framework of the bisymmetric model proposed by \citet{SS07}, and investigate the dependence of bar-induced velocity perturbations (i.e., noncircular motions) and rotation curves on bar strength. We also perform mock observations of the numerical data to assess how well the {\tt XookSuut} code decomposes the observed \ac{LOS} velocities into radial and tangential components. Our main findings can be summarized as follows:

\begin{enumerate}

\item Stars moving on bar-supporting $x_1$ orbits exhibit characteristic sawtooth-like patterns in radial velocity and arch-like patterns in tangential velocity, both with a periodicity of $\pi$ in the azimuthal direction. These features arise naturally from the elongation of $x_1$ orbits along the bar, with their amplitudes increasing as the bar becomes stronger (see \autoref{eq:x1fits}).

\item 
We classify the orbits of particles in the bar region into three groups based on the ratio of peak amplitudes $A_x$ and $A_y$, measured parallel and perpendicular to the bar, respectively, as well as the time-averaged dimensionless angular momentum $\mathcal{L}$ (see \autoref{eq:L}): $x_1$-like orbits for $A_x/A_y\geq1.5$ and $\mathcal{L}\geq0.25$; box orbits for $A_x/A_y\geq1.5$ and $\mathcal{L}<0.25$, and disk orbits for $A_x/A_y<1.5$. While particles on $x_1$-like orbits exhibit the characteristic sawtooth-like and arch-shaped patterns in their velocity distributions, those on box and disk orbits display no significant azimuthal variation.

\item 
Particles on box and disk orbits act to blur the distinctive sawtooth-like and arch-shaped velocity patterns of $x_1$-like orbits, thereby rendering the overall velocity distributions broadly consistent with the predictions of the bisymmetric model of \citet{SS07}. The bisymmetric sinusoidal patterns (see \autoref{eq:bisymmetric0}), characterized by a $45^\circ$ phase offset between the radial and tangential components, capture the angular dependence of stellar velocities in barred galaxies reasonably well, albeit only in an averaged sense. The associated velocity dispersions, on the order of $\sim 50$--$100\kms$, are comparable in magnitude to the mean velocities. (see \cref{fig:bisymmetric_C10,fig:bisymmetric_C20}).

\item 
The amplitude $A_\text{bis}(R)$ of the bisymmetric perturbations, as defined in \cref{eq:Abis}, reaches its maximum at a radius $R_\text{max} \sim 0.5 R_b$,
in agreement with the observational results of \citet{Kim24}. The mean value of $A_\text{bis}$ and its fractional counterpart, $f_\text{NC}$, relative to the unperturbed rotational velocity, increase with bar strength. The linear correlations of $A_\text{bis}$ and $f_\text{NC}$ with $A_2/A_0$, as shown in \cref{eq:Abisfit,eq:fNCfit}, respectively, support the conclusion of \citet{Kim24} that bars are the primary drivers of noncircular motions in barred galaxies.

\item The presence of a bar enhances the radial pressure gradient force arising from random stellar motions, resulting in a noticeable reduction in the galaxy’s rotation curve. This reduction becomes more pronounced with increasing bar strength. As a result, galaxies with stronger bars tend to exhibit shallower rotation curves within the bar region, in broad agreement with the combined observational findings of \citet{LopezCoba22} and \citet{Kim24}.

\item 
Synthetic observations of our model galaxies show that decomposing the \ac{LOS} velocities into radial and tangential components using {\tt XookSuut}, based on the bisymmetric model, is reasonably accurate when the bar \ac{PA} is supplied as an input parameter, which can be readily inferred from photometric data. However, if the bar \ac{PA} is treated as a free parameter, the resulting amplitudes of the decomposed velocities may be unreliable.
\end{enumerate}

\section*{Acknowledgments}
We are grateful to the referee for constructive comments. This work was supported by the National Research Foundation of Korea (NRF) grant funded by the Korea government (MSIT) (RS-2025-00517264). The work of D.J. was supported by Basic Science Research Program through the NRF funded by the Ministry of Education (RS-2023-00273275). T.K. acknowledges support from the Basic Science Research Program through the National Research Foundation of Korea (NRF) funded by the Ministry of Education (No. RS-2023-00240212). Computational resources for this project were provided by the Supercomputing Center/Korea Institute of Science and Technology Information with supercomputing resources including technical support (KSC-2023-CRE-0175).

\bibliography{biblist}{}

\begin{thebibliography}{}
\expandafter\ifx\csname natexlab\endcsname\relax\def\natexlab#1{#1}\fi
\providecommand{\url}[1]{\href{#1}{#1}}
\providecommand{\dodoi}[1]{doi:~\href{http://doi.org/#1}{\nolinkurl{#1}}}
\providecommand{\doeprint}[1]{\href{http://ascl.net/#1}{\nolinkurl{http://ascl.net/#1}}}
\providecommand{\doarXiv}[1]{\href{https://arxiv.org/abs/#1}{\nolinkurl{https://arxiv.org/abs/#1}}}

\bibitem[{{Abraham} {et~al.}(1999){Abraham}, {Merrifield}, {Ellis}, {Tanvir}, \& {Brinchmann}}]{Abraham99}
{Abraham}, R.~G., {Merrifield}, M.~R., {Ellis}, R.~S., {Tanvir}, N.~R., \& {Brinchmann}, J. 1999, \mnras, 308, 569, \dodoi{10.1046/j.1365-8711.1999.02766.x}

\bibitem[{{Aguerri} {et~al.}(2009){Aguerri}, {M{\'e}ndez-Abreu}, \& {Corsini}}]{Aguerri09}
{Aguerri}, J.~A.~L., {M{\'e}ndez-Abreu}, J., \& {Corsini}, E.~M. 2009, \aap, 495, 491, \dodoi{10.1051/0004-6361:200810931}

\bibitem[{{Algorry} {et~al.}(2017){Algorry}, {Navarro}, {Abadi}, {Sales}, {Bower}, {Crain}, {Dalla Vecchia}, {Frenk}, {Schaller}, {Schaye}, \& {Theuns}}]{Algorry17}
{Algorry}, D.~G., {Navarro}, J.~F., {Abadi}, M.~G., {et~al.} 2017, \mnras, 469, 1054, \dodoi{10.1093/mnras/stx1008}

\bibitem[{{Amvrosiadis} {et~al.}(2025){Amvrosiadis}, {Lange}, {Nightingale}, {He}, {Frenk}, {Oman}, {Smail}, {Swinbank}, {Fragkoudi}, {Gadotti}, {Cole}, {Borsato}, {Robertson}, {Massey}, {Cao}, \& {Li}}]{Amvrosiadis24}
{Amvrosiadis}, A., {Lange}, S., {Nightingale}, J.~W., {et~al.} 2025, \mnras, 537, 1163, \dodoi{10.1093/mnras/staf048}

\bibitem[{{Athanassoula}(2013)}]{Athanassoula13}
{Athanassoula}, E. 2013, in Secular Evolution of Galaxies, ed. J.~{Falc{\'o}n-Barroso} \& J.~H. {Knapen}, 305, \dodoi{10.48550/arXiv.1211.6752}

\bibitem[{{Beckman} {et~al.}(2004){Beckman}, {Zurita}, \& {Vega Beltr{\'a}n}}]{Beckman04}
{Beckman}, J.~E., {Zurita}, A., \& {Vega Beltr{\'a}n}, J.~C. 2004, in Lecture Notes and Essays in Astrophysics, ed. A.~{Ulla} \& M.~{Manteiga}, Vol.~1, 43--62

\bibitem[{{Binney} \& {Spergel}(1982)}]{Binney82}
{Binney}, J., \& {Spergel}, D. 1982, \apj, 252, 308, \dodoi{10.1086/159559}

\bibitem[{{Binney} \& {Spergel}(1984)}]{Binney84}
---. 1984, \mnras, 206, 159, \dodoi{10.1093/mnras/206.1.159}

\bibitem[{{Binney} \& {Tremaine}(2008)}]{BT08}
{Binney}, J., \& {Tremaine}, S. 2008, {Galactic Dynamics: Second Edition}

\bibitem[{{Bland-Hawthorn} \& {Gerhard}(2016)}]{Bland-Hawthorn16}
{Bland-Hawthorn}, J., \& {Gerhard}, O. 2016, \araa, 54, 529, \dodoi{10.1146/annurev-astro-081915-023441}

\bibitem[{{Bosma}(1978)}]{Bosma78}
{Bosma}, A. 1978, PhD thesis, University of Groningen, Netherlands

\bibitem[{{Bureau} \& {Athanassoula}(2005)}]{Bureau05}
{Bureau}, M., \& {Athanassoula}, E. 2005, \apj, 626, 159, \dodoi{10.1086/430056}

\bibitem[{{Buta} {et~al.}(2015){Buta}, {Sheth}, {Athanassoula}, {Bosma}, {Knapen}, {Laurikainen}, {Salo}, {Elmegreen}, {Ho}, {Zaritsky}, {Courtois}, {Hinz}, {Mu{\~n}oz-Mateos}, {Kim}, {Regan}, {Gadotti}, {Gil de Paz}, {Laine}, {Men{\'e}ndez-Delmestre}, {Comer{\'o}n}, {Erroz Ferrer}, {Seibert}, {Mizusawa}, {Holwerda}, \& {Madore}}]{Buta15}
{Buta}, R.~J., {Sheth}, K., {Athanassoula}, E., {et~al.} 2015, \apjs, 217, 32, \dodoi{10.1088/0067-0049/217/2/32}

\bibitem[{{Cameron} {et~al.}(2010){Cameron}, {Carollo}, {Oesch}, {Aller}, {Bschorr}, {Cerulo}, {Aussel}, {Capak}, {Le Floc'h}, {Ilbert}, {Kneib}, {Koekemoer}, {Leauthaud}, {Lilly}, {Massey}, {McCracken}, {Rhodes}, {Salvato}, {Sanders}, {Scoville}, {Sheth}, {Taniguchi}, \& {Thompson}}]{Cameron10}
{Cameron}, E., {Carollo}, C.~M., {Oesch}, P., {et~al.} 2010, \mnras, 409, 346, \dodoi{10.1111/j.1365-2966.2010.17314.x}

\bibitem[{{Chaves-Velasquez} {et~al.}(2017){Chaves-Velasquez}, {Patsis}, {Puerari}, {Skokos}, \& {Manos}}]{Chaves17}
{Chaves-Velasquez}, L., {Patsis}, P.~A., {Puerari}, I., {Skokos}, C., \& {Manos}, T. 2017, \apj, 850, 145, \dodoi{10.3847/1538-4357/aa961a}

\bibitem[{{Chung} \& {Bureau}(2004)}]{Chung04}
{Chung}, A., \& {Bureau}, M. 2004, \aj, 127, 3192, \dodoi{10.1086/420988}

\bibitem[{{Costantin} {et~al.}(2023){Costantin}, {P{\'e}rez-Gonz{\'a}lez}, {Guo}, {Buttitta}, {Jogee}, {Bagley}, {Barro}, {Kartaltepe}, {Koekemoer}, {Cabello}, {Corsini}, {M{\'e}ndez-Abreu}, {de la Vega}, {Iyer}, {Bisigello}, {Cheng}, {Morelli}, {Arrabal Haro}, {Buitrago}, {Cooper}, {Dekel}, {Dickinson}, {Finkelstein}, {Giavalisco}, {Holwerda}, {Huertas-Company}, {Lucas}, {Papovich}, {Pirzkal}, {Seill{\'e}}, {Vega-Ferrero}, {Wuyts}, \& {Yung}}]{Costantin23}
{Costantin}, L., {P{\'e}rez-Gonz{\'a}lez}, P.~G., {Guo}, Y., {et~al.} 2023, \nat, 623, 499, \dodoi{10.1038/s41586-023-06636-x}

\bibitem[{{de Vaucouleurs}(1963)}]{deVaucouleurs63}
{de Vaucouleurs}, G. 1963, \apj, 138, 934, \dodoi{10.1086/147696}

\bibitem[{{de Vaucouleurs} {et~al.}(1991){de Vaucouleurs}, {de Vaucouleurs}, {Corwin}, {Buta}, {Paturel}, \& {Fouque}}]{deVaucouleurs91}
{de Vaucouleurs}, G., {de Vaucouleurs}, A., {Corwin}, Jr., H.~G., {et~al.} 1991, {Third Reference Catalogue of Bright Galaxies}

\bibitem[{{D{\'\i}az-Garc{\'\i}a} {et~al.}(2019){D{\'\i}az-Garc{\'\i}a}, {Salo}, {Knapen}, \& {Herrera-Endoqui}}]{Diaz19}
{D{\'\i}az-Garc{\'\i}a}, S., {Salo}, H., {Knapen}, J.~H., \& {Herrera-Endoqui}, M. 2019, \aap, 631, A94, \dodoi{10.1051/0004-6361/201936000}

\bibitem[{{D{\'\i}az-Garc{\'\i}a} {et~al.}(2016){D{\'\i}az-Garc{\'\i}a}, {Salo}, {Laurikainen}, \& {Herrera-Endoqui}}]{Diaz16}
{D{\'\i}az-Garc{\'\i}a}, S., {Salo}, H., {Laurikainen}, E., \& {Herrera-Endoqui}, M. 2016, \aap, 587, A160, \dodoi{10.1051/0004-6361/201526161}

\bibitem[{{DiGiorgio Zanger} {et~al.}(2024){DiGiorgio Zanger}, {Westfall}, {Bundy}, {Drory}, {Bershady}, {Campbell}, {Weijmans}, {Masters}, {Stark}, \& {Law}}]{Zanger24}
{DiGiorgio Zanger}, B., {Westfall}, K.~B., {Bundy}, K., {et~al.} 2024, \apj, 973, 116, \dodoi{10.3847/1538-4357/ad6606}

\bibitem[{{Eskridge} {et~al.}(2000){Eskridge}, {Frogel}, {Pogge}, {Quillen}, {Davies}, {DePoy}, {Houdashelt}, {Kuchinski}, {Ram{\'\i}rez}, {Sellgren}, {Terndrup}, \& {Tiede}}]{Eskridge00}
{Eskridge}, P.~B., {Frogel}, J.~A., {Pogge}, R.~W., {et~al.} 2000, \aj, 119, 536, \dodoi{10.1086/301203}

\bibitem[{{Fathi} {et~al.}(2005){Fathi}, {van de Ven}, {Peletier}, {Emsellem}, {Falc{\'o}n-Barroso}, {Cappellari}, \& {de Zeeuw}}]{Fathi05}
{Fathi}, K., {van de Ven}, G., {Peletier}, R.~F., {et~al.} 2005, \mnras, 364, 773, \dodoi{10.1111/j.1365-2966.2005.09648.x}

\bibitem[{{Gajda} {et~al.}(2016){Gajda}, {{\L}okas}, \& {Athanassoula}}]{Gajda16}
{Gajda}, G., {{\L}okas}, E.~L., \& {Athanassoula}, E. 2016, \apj, 830, 108, \dodoi{10.3847/0004-637X/830/2/108}

\bibitem[{{G{\'e}ron} {et~al.}(2025){G{\'e}ron}, {Smethurst}, {Dickinson}, {Fortson}, {Garland}, {Kruk}, {Lintott}, {Makechemu}, {Mantha}, {Masters}, {O'Ryan}, {Roberts}, {Simmons}, {Walmsley}, {Calabr{\`o}}, {Chiba}, {Costantin}, {Drout}, {Fragkoudi}, {Guo}, {Holwerda}, {Jogee}, {Koekemoer}, {Lucas}, \& {Pacucci}}]{Geron25}
{G{\'e}ron}, T., {Smethurst}, R.~J., {Dickinson}, H., {et~al.} 2025, \apj, 987, 74, \dodoi{10.3847/1538-4357/add7d0}

\bibitem[{{Ghosh} {et~al.}(2025){Ghosh}, {Kalda}, {Di Matteo}, {Green}, {Khoperskov}, {Katz}, \& {Haywood}}]{Ghosh25}
{Ghosh}, S., {Kalda}, T., {Di Matteo}, P., {et~al.} 2025, arXiv e-prints, arXiv:2504.06352, \dodoi{10.48550/arXiv.2504.06352}

\bibitem[{{Guo} {et~al.}(2023){Guo}, {Jogee}, {Finkelstein}, {Chen}, {Wise}, {Bagley}, {Barro}, {Wuyts}, {Kocevski}, {Kartaltepe}, {McGrath}, {Ferguson}, {Mobasher}, {Giavalisco}, {Lucas}, {Zavala}, {Lotz}, {Grogin}, {Huertas-Company}, {Vega-Ferrero}, {Hathi}, {Arrabal Haro}, {Dickinson}, {Koekemoer}, {Papovich}, {Pirzkal}, {Yung}, {Backhaus}, {Bell}, {Calabr{\`o}}, {Cleri}, {Coogan}, {Cooper}, {Costantin}, {Croton}, {Davis}, {Dekel}, {Franco}, {Gardner}, {Holwerda}, {Hutchison}, {Pandya}, {P{\'e}rez-Gonz{\'a}lez}, {Ravindranath}, {Rose}, {Trump}, {de la Vega}, \& {Wang}}]{Guo23}
{Guo}, Y., {Jogee}, S., {Finkelstein}, S.~L., {et~al.} 2023, \apjl, 945, L10, \dodoi{10.3847/2041-8213/acacfb}

\bibitem[{{Hallenbeck} {et~al.}(2014){Hallenbeck}, {Huang}, {Spekkens}, {Haynes}, {Giovanelli}, {Adams}, {Brinchmann}, {Chengalur}, {Hunt}, {Masters}, \& {Saintonge}}]{Hallenbeck14}
{Hallenbeck}, G., {Huang}, S., {Spekkens}, K., {et~al.} 2014, \aj, 148, 69, \dodoi{10.1088/0004-6256/148/4/69}

\bibitem[{{Helmi}(2020)}]{Helmi20}
{Helmi}, A. 2020, \araa, 58, 205, \dodoi{10.1146/annurev-astro-032620-021917}

\bibitem[{{Hernquist}(1990)}]{Hernquist90}
{Hernquist}, L. 1990, \apj, 356, 359, \dodoi{10.1086/168845}

\bibitem[{{Hodge} {et~al.}(2019){Hodge}, {Smail}, {Walter}, {da Cunha}, {Swinbank}, {Rybak}, {Venemans}, {Brandt}, {Calistro Rivera}, {Chapman}, {Chen}, {Cox}, {Dannerbauer}, {Decarli}, {Greve}, {Knudsen}, {Menten}, {Schinnerer}, {Simpson}, {van der Werf}, {Wardlow}, \& {Weiss}}]{Hodge19}
{Hodge}, J.~A., {Smail}, I., {Walter}, F., {et~al.} 2019, \apj, 876, 130, \dodoi{10.3847/1538-4357/ab1846}

\bibitem[{{Hogarth} {et~al.}(2024){Hogarth}, {Saintonge}, {Davis}, {Ellison}, {Lin}, {L{\'o}pez-Cob{\'a}}, {Pan}, \& {Thorp}}]{Hogarth24}
{Hogarth}, L.~M., {Saintonge}, A., {Davis}, T.~A., {et~al.} 2024, \mnras, 528, 6768, \dodoi{10.1093/mnras/stae377}

\bibitem[{{Holmes} {et~al.}(2015){Holmes}, {Spekkens}, {S{\'a}nchez}, {Walcher}, {Garc{\'\i}a-Benito}, {Mast}, {Cortijo-Ferrero}, {Kalinova}, {Marino}, {Mendez-Abreu}, \& {Barrera-Ballesteros}}]{Holmes15}
{Holmes}, L., {Spekkens}, K., {S{\'a}nchez}, S.~F., {et~al.} 2015, \mnras, 451, 4397, \dodoi{10.1093/mnras/stv1254}

\bibitem[{{Huertas-Company} {et~al.}(2025){Huertas-Company}, {Shuntov}, {Dong}, {Walmsley}, {Ilbert}, {McCracken}, {Akins}, {Allen}, {Casey}, {Costantin}, {Daddi}, {Dekel}, {Franco}, {Garland}, {G{\'e}ron}, {Gozaliasl}, {Hirschmann}, {Kartaltepe}, {Koekemoer}, {Lintott}, {Liu}, {Lucas}, {Masters}, {Pacucci}, {Paquereau}, {P'erez-Gonz'alez}, {Rhodes}, {Robertson}, {Simmons}, {Smethurst}, {Toft}, \& {Yang}}]{Huertas25}
{Huertas-Company}, M., {Shuntov}, M., {Dong}, Y., {et~al.} 2025, arXiv e-prints, arXiv:2502.03532, \dodoi{10.48550/arXiv.2502.03532}

\bibitem[{{Huntley}(1978)}]{Huntley78}
{Huntley}, J.~M. 1978, \apjl, 225, L101, \dodoi{10.1086/182803}

\bibitem[{{Jang} \& {Kim}(2023)}]{JK23}
{Jang}, D., \& {Kim}, W.-T. 2023, \apj, 942, 106, \dodoi{10.3847/1538-4357/aca7bc}

\bibitem[{{Kim} {et~al.}(2024){Kim}, {Gadotti}, {Lee}, {L{\'o}pez-Cob{\'a}}, {Kim}, {Kim}, \& {Park}}]{Kim24}
{Kim}, T., {Gadotti}, D.~A., {Lee}, Y.~H., {et~al.} 2024, \apj, 976, 220, \dodoi{10.3847/1538-4357/ad8573}

\bibitem[{{Knapen} {et~al.}(2000){Knapen}, {Shlosman}, \& {Peletier}}]{Knapen00}
{Knapen}, J.~H., {Shlosman}, I., \& {Peletier}, R.~F. 2000, \apj, 529, 93, \dodoi{10.1086/308266}

\bibitem[{{Kormendy}(1983)}]{Kormendy83}
{Kormendy}, J. 1983, \apj, 275, 529, \dodoi{10.1086/161552}

\bibitem[{{Laskar}(1990)}]{Laskar90}
{Laskar}, J. 1990, \icarus, 88, 266, \dodoi{10.1016/0019-1035(90)90084-M}

\bibitem[{{Laskar}(1993)}]{Laskar93}
---. 1993, CeMDA, 56, 191, \dodoi{10.1007/BF00699731}

\bibitem[{{Laurikainen} {et~al.}(2004){Laurikainen}, {Salo}, \& {Buta}}]{Laurikainen04}
{Laurikainen}, E., {Salo}, H., \& {Buta}, R. 2004, \apj, 607, 103, \dodoi{10.1086/383462}

\bibitem[{{Le Conte} {et~al.}(2024){Le Conte}, {Gadotti}, {Ferreira}, {Conselice}, {de S{\'a}-Freitas}, {Kim}, {Neumann}, {Fragkoudi}, {Athanassoula}, \& {Adams}}]{LeConte24}
{Le Conte}, Z.~A., {Gadotti}, D.~A., {Ferreira}, L., {et~al.} 2024, \mnras, 530, 1984, \dodoi{10.1093/mnras/stae921}

\bibitem[{{Liu} {et~al.}(2025){Liu}, {Li}, \& {Shen}}]{Liu25}
{Liu}, J., {Li}, Z., \& {Shen}, J. 2025, \apj, 980, 146, \dodoi{10.3847/1538-4357/adabe0}

\bibitem[{{{\L}okas}(2025)}]{Lokas25}
{{\L}okas}, E.~L. 2025, \aap, 699, A234, \dodoi{10.1051/0004-6361/202553817}

\bibitem[{{L{\'o}pez-Cob{\'a}} {et~al.}(2024{\natexlab{a}}){L{\'o}pez-Cob{\'a}}, {Lin}, {Neumann}, \& {Bershady}}]{LopezCoba24a}
{L{\'o}pez-Cob{\'a}}, C., {Lin}, L., {Neumann}, J., \& {Bershady}, M.~A. 2024{\natexlab{a}}, \apj, 975, 182, \dodoi{10.3847/1538-4357/ad7b12}

\bibitem[{{L{\'o}pez-Cob{\'a}} {et~al.}(2024{\natexlab{b}}){L{\'o}pez-Cob{\'a}}, {Lin}, \& {S{\'a}nchez}}]{LopezCoba24b}
{L{\'o}pez-Cob{\'a}}, C., {Lin}, L., \& {S{\'a}nchez}, S.~F. 2024{\natexlab{b}}, \apj, 962, 65, \dodoi{10.3847/1538-4357/ad152a}

\bibitem[{{L{\'o}pez-Cob{\'a}} {et~al.}(2024{\natexlab{c}}){L{\'o}pez-Cob{\'a}}, {Lin}, \& {S{\'a}nchez}}]{LopezCoba24c}
---. 2024{\natexlab{c}}, \rmxaa, 60, 19, \dodoi{10.22201/ia.01851101p.2024.60.01.03}

\bibitem[{{L{\'o}pez-Cob{\'a}} {et~al.}(2022){L{\'o}pez-Cob{\'a}}, {S{\'a}nchez}, {Lin}, {Anderson}, {Lin}, {Cruz-Gonz{\'a}lez}, {Galbany}, \& {Barrera-Ballesteros}}]{LopezCoba22}
{L{\'o}pez-Cob{\'a}}, C., {S{\'a}nchez}, S.~F., {Lin}, L., {et~al.} 2022, \apj, 939, 40, \dodoi{10.3847/1538-4357/ac937b}

\bibitem[{{L{\'o}pez-Cob{\'a}} {et~al.}(2020){L{\'o}pez-Cob{\'a}}, {S{\'a}nchez}, {Anderson}, {Cruz-Gonz{\'a}lez}, {Galbany}, {Ruiz-Lara}, {Barrera-Ballesteros}, {Prieto}, \& {Kuncarayakti}}]{LopezCoba20}
{L{\'o}pez-Cob{\'a}}, C., {S{\'a}nchez}, S.~F., {Anderson}, J.~P., {et~al.} 2020, \aj, 159, 167, \dodoi{10.3847/1538-3881/ab7848}

\bibitem[{{Marinova} \& {Jogee}(2007)}]{Marinova07}
{Marinova}, I., \& {Jogee}, S. 2007, \apj, 659, 1176, \dodoi{10.1086/512355}

\bibitem[{{M{\'a}rquez} {et~al.}(2002){M{\'a}rquez}, {Masegosa}, {Moles}, {Varela}, {Bettoni}, \& {Galletta}}]{Marquez02}
{M{\'a}rquez}, I., {Masegosa}, J., {Moles}, M., {et~al.} 2002, \aap, 393, 389, \dodoi{10.1051/0004-6361:20021036}

\bibitem[{{Masters} {et~al.}(2011){Masters}, {Nichol}, {Hoyle}, {Lintott}, {Bamford}, {Edmondson}, {Fortson}, {Keel}, {Schawinski}, {Smith}, \& {Thomas}}]{Masters11}
{Masters}, K.~L., {Nichol}, R.~C., {Hoyle}, B., {et~al.} 2011, \mnras, 411, 2026, \dodoi{10.1111/j.1365-2966.2010.17834.x}

\bibitem[{{Melvin} {et~al.}(2014){Melvin}, {Masters}, {Lintott}, {Nichol}, {Simmons}, {Bamford}, {Casteels}, {Cheung}, {Edmondson}, {Fortson}, {Schawinski}, {Skibba}, {Smith}, \& {Willett}}]{Melvin14}
{Melvin}, T., {Masters}, K., {Lintott}, C., {et~al.} 2014, \mnras, 438, 2882, \dodoi{10.1093/mnras/stt2397}

\bibitem[{{M{\'e}ndez-Abreu} {et~al.}(2010){M{\'e}ndez-Abreu}, {S{\'a}nchez-Janssen}, \& {Aguerri}}]{Men10}
{M{\'e}ndez-Abreu}, J., {S{\'a}nchez-Janssen}, R., \& {Aguerri}, J.~A.~L. 2010, \apjl, 711, L61, \dodoi{10.1088/2041-8205/711/2/L61}

\bibitem[{{M{\'e}ndez-Abreu} {et~al.}(2012){M{\'e}ndez-Abreu}, {S{\'a}nchez-Janssen}, {Aguerri}, {Corsini}, \& {Zarattini}}]{Men12}
{M{\'e}ndez-Abreu}, J., {S{\'a}nchez-Janssen}, R., {Aguerri}, J.~A.~L., {Corsini}, E.~M., \& {Zarattini}, S. 2012, \apjl, 761, L6, \dodoi{10.1088/2041-8205/761/1/L6}

\bibitem[{{Men{\'e}ndez-Delmestre} {et~al.}(2007){Men{\'e}ndez-Delmestre}, {Sheth}, {Schinnerer}, {Jarrett}, \& {Scoville}}]{Men07}
{Men{\'e}ndez-Delmestre}, K., {Sheth}, K., {Schinnerer}, E., {Jarrett}, T.~H., \& {Scoville}, N.~Z. 2007, \apj, 657, 790, \dodoi{10.1086/511025}

\bibitem[{{Minchev} \& {Famaey}(2010)}]{Minchev10}
{Minchev}, I., \& {Famaey}, B. 2010, \apj, 722, 112, \dodoi{10.1088/0004-637X/722/1/112}

\bibitem[{{Molaeinezhad} {et~al.}(2016){Molaeinezhad}, {Falc{\'o}n-Barroso}, {Mart{\'\i}nez-Valpuesta}, {Khosroshahi}, {Balcells}, \& {Peletier}}]{Molaeinezhad16}
{Molaeinezhad}, A., {Falc{\'o}n-Barroso}, J., {Mart{\'\i}nez-Valpuesta}, I., {et~al.} 2016, \mnras, 456, 692, \dodoi{10.1093/mnras/stv2697}

\bibitem[{{Nair} \& {Abraham}(2010)}]{Nair10}
{Nair}, P.~B., \& {Abraham}, R.~G. 2010, \apjs, 186, 427, \dodoi{10.1088/0067-0049/186/2/427}

\bibitem[{{Papaphilippou} \& {Laskar}(1996)}]{Papaphilippou96}
{Papaphilippou}, Y., \& {Laskar}, J. 1996, \aap, 307, 427

\bibitem[{{Papaphilippou} \& {Laskar}(1998)}]{Papaphilippou98}
---. 1998, \aap, 329, 451

\bibitem[{{Parul} {et~al.}(2020){Parul}, {Smirnov}, \& {Sotnikova}}]{Parul20}
{Parul}, H.~D., {Smirnov}, A.~A., \& {Sotnikova}, N.~Y. 2020, \apj, 895, 12, \dodoi{10.3847/1538-4357/ab76ce}

\bibitem[{{Patsis} \& {Harsoula}(2018)}]{Patsis18}
{Patsis}, P.~A., \& {Harsoula}, M. 2018, \aap, 612, A114, \dodoi{10.1051/0004-6361/201731114}

\bibitem[{{Pejch} {et~al.}(2023){Pejch}, {Morozov}, \& {Khoperskov}}]{Pejch23}
{Pejch}, M., {Morozov}, A., \& {Khoperskov}, A.~V. 2023, Mathematical Physics and Computer Simulation, 3, 91

\bibitem[{{Pence}(1981)}]{Pence81}
{Pence}, W.~D. 1981, \apj, 247, 473, \dodoi{10.1086/159056}

\bibitem[{{Pence} \& {Blackman}(1984)}]{Pence84}
{Pence}, W.~D., \& {Blackman}, C.~P. 1984, \mnras, 207, 9, \dodoi{10.1093/mnras/207.1.9}

\bibitem[{{Peterson} \& {Huntley}(1980)}]{Peterson80}
{Peterson}, C.~J., \& {Huntley}, J.~M. 1980, \apj, 242, 913, \dodoi{10.1086/158525}

\bibitem[{{Peterson} {et~al.}(1978){Peterson}, {Rubin}, {Ford}, \& {Thonnard}}]{Peterson78}
{Peterson}, C.~J., {Rubin}, V.~C., {Ford}, Jr., W.~K., \& {Thonnard}, N. 1978, \apj, 219, 31, \dodoi{10.1086/155752}

\bibitem[{{Pinna} {et~al.}(2018){Pinna}, {Falc{\'o}n-Barroso}, {Martig}, {Mart{\'\i}nez-Valpuesta}, {M{\'e}ndez-Abreu}, {van de Ven}, {Leaman}, \& {Lyubenova}}]{Pinna18}
{Pinna}, F., {Falc{\'o}n-Barroso}, J., {Martig}, M., {et~al.} 2018, \mnras, 475, 2697, \dodoi{10.1093/mnras/stx3331}

\bibitem[{{Saburova} {et~al.}(2017){Saburova}, {Katkov}, {Khoperskov}, {Zasov}, \& {Uklein}}]{Saburova17}
{Saburova}, A.~S., {Katkov}, I.~Y., {Khoperskov}, S.~A., {Zasov}, A.~V., \& {Uklein}, R.~I. 2017, \mnras, 470, 20, \dodoi{10.1093/mnras/stx1200}

\bibitem[{{Seidel} {et~al.}(2015){Seidel}, {Falc{\'o}n-Barroso}, {Mart{\'\i}nez-Valpuesta}, {D{\'\i}az-Garc{\'\i}a}, {Laurikainen}, {Salo}, \& {Knapen}}]{Seidel15}
{Seidel}, M.~K., {Falc{\'o}n-Barroso}, J., {Mart{\'\i}nez-Valpuesta}, I., {et~al.} 2015, \mnras, 451, 936, \dodoi{10.1093/mnras/stv969}

\bibitem[{{Sellwood} \& {Spekkens}(2015)}]{SS15}
{Sellwood}, J.~A., \& {Spekkens}, K. 2015, arXiv e-prints, arXiv:1509.07120, \dodoi{10.48550/arXiv.1509.07120}

\bibitem[{{Sellwood} \& {Wilkinson}(1993)}]{SW93}
{Sellwood}, J.~A., \& {Wilkinson}, A. 1993, Reports on Progress in Physics, 56, 173, \dodoi{10.1088/0034-4885/56/2/001}

\bibitem[{{Shen} {et~al.}(2010){Shen}, {Rich}, {Kormendy}, {Howard}, {De Propris}, \& {Kunder}}]{Shen10}
{Shen}, J., {Rich}, R.~M., {Kormendy}, J., {et~al.} 2010, \apjl, 720, L72, \dodoi{10.1088/2041-8205/720/1/L72}

\bibitem[{{Shen} \& {Sellwood}(2004)}]{SS04}
{Shen}, J., \& {Sellwood}, J.~A. 2004, \apj, 604, 614, \dodoi{10.1086/382124}

\bibitem[{{Sheth} {et~al.}(2008){Sheth}, {Elmegreen}, {Elmegreen}, {Capak}, {Abraham}, {Athanassoula}, {Ellis}, {Mobasher}, {Salvato}, {Schinnerer}, {Scoville}, {Spalsbury}, {Strubbe}, {Carollo}, {Rich}, \& {West}}]{Sheth08}
{Sheth}, K., {Elmegreen}, D.~M., {Elmegreen}, B.~G., {et~al.} 2008, \apj, 675, 1141, \dodoi{10.1086/524980}

\bibitem[{{Smail} {et~al.}(2023){Smail}, {Dudzevi{\v{c}}i{\={u}}t{\.{e}}}, {Gurwell}, {Fazio}, {Willner}, {Swinbank}, {Arumugam}, {Summers}, {Cohen}, {Jansen}, {Windhorst}, {Meena}, {Zitrin}, {Keel}, {Cheng}, {Coe}, {Conselice}, {D'Silva}, {Driver}, {Frye}, {Grogin}, {Koekemoer}, {Marshall}, {Nonino}, {Pirzkal}, {Robotham}, {Rutkowski}, {Ryan}, {Tompkins}, {Willmer}, {Yan}, {Broadhurst}, {Diego}, {Kamieneski}, \& {Yun}}]{Smail23}
{Smail}, I., {Dudzevi{\v{c}}i{\={u}}t{\.{e}}}, U., {Gurwell}, M., {et~al.} 2023, \apj, 958, 36, \dodoi{10.3847/1538-4357/acf931}

\bibitem[{{Smirnov} {et~al.}(2021){Smirnov}, {Tikhonenko}, \& {Sotnikova}}]{Smirnov21}
{Smirnov}, A.~A., {Tikhonenko}, I.~S., \& {Sotnikova}, N.~Y. 2021, \mnras, 502, 4689, \dodoi{10.1093/mnras/stab327}

\bibitem[{{Spekkens} \& {Sellwood}(2007)}]{SS07}
{Spekkens}, K., \& {Sellwood}, J.~A. 2007, \apj, 664, 204, \dodoi{10.1086/518471}

\bibitem[{{Springel} {et~al.}(2021){Springel}, {Pakmor}, {Zier}, \& {Reinecke}}]{Springel21}
{Springel}, V., {Pakmor}, R., {Zier}, O., \& {Reinecke}, M. 2021, \mnras, 506, 2871, \dodoi{10.1093/mnras/stab1855}

\bibitem[{{Stark} {et~al.}(2018){Stark}, {Bundy}, {Westfall}, {Bershady}, {Weijmans}, {Masters}, {Kruk}, {Brinchmann}, {Soler}, {Abraham}, {Cheung}, {Bizyaev}, {Drory}, {Lopes}, \& {Law}}]{Stark18}
{Stark}, D.~V., {Bundy}, K.~A., {Westfall}, K., {et~al.} 2018, \mnras, 480, 2217, \dodoi{10.1093/mnras/sty1991}

\bibitem[{{Tsukui} {et~al.}(2024){Tsukui}, {Wisnioski}, {Bland-Hawthorn}, {Mai}, {Iguchi}, {Baba}, \& {Freeman}}]{Tsukui24}
{Tsukui}, T., {Wisnioski}, E., {Bland-Hawthorn}, J., {et~al.} 2024, \mnras, 527, 8941, \dodoi{10.1093/mnras/stad3588}

\bibitem[{{Valencia-Enr{\'\i}quez} {et~al.}(2023){Valencia-Enr{\'\i}quez}, {Puerari}, \& {Chaves-Velasquez}}]{Valencia23}
{Valencia-Enr{\'\i}quez}, D., {Puerari}, I., \& {Chaves-Velasquez}, L. 2023, \mnras, 525, 3162, \dodoi{10.1093/mnras/stad2437}

\bibitem[{{Valluri} {et~al.}(2010){Valluri}, {Debattista}, {Quinn}, \& {Moore}}]{Valluri10}
{Valluri}, M., {Debattista}, V.~P., {Quinn}, T., \& {Moore}, B. 2010, \mnras, 403, 525, \dodoi{10.1111/j.1365-2966.2009.16192.x}

\bibitem[{{Valluri} \& {Merritt}(1998)}]{Valluri98}
{Valluri}, M., \& {Merritt}, D. 1998, \apj, 506, 686, \dodoi{10.1086/306269}

\bibitem[{{Valluri} {et~al.}(2016){Valluri}, {Shen}, {Abbott}, \& {Debattista}}]{Valluri16}
{Valluri}, M., {Shen}, J., {Abbott}, C., \& {Debattista}, V.~P. 2016, \apj, 818, 141, \dodoi{10.3847/0004-637X/818/2/141}

\bibitem[{{Voglis} {et~al.}(2007){Voglis}, {Harsoula}, \& {Contopoulos}}]{Voglis07}
{Voglis}, N., {Harsoula}, M., \& {Contopoulos}, G. 2007, \mnras, 381, 757, \dodoi{10.1111/j.1365-2966.2007.12263.x}

\bibitem[{{Walo-Mart{\'\i}n} {et~al.}(2022){Walo-Mart{\'\i}n}, {Pinna}, {Grand}, {P{\'e}rez}, {Falc{\'o}n-Barroso}, {Fragkoudi}, \& {Martig}}]{WaloMartin22}
{Walo-Mart{\'\i}n}, D., {Pinna}, F., {Grand}, R. J.~J., {et~al.} 2022, \mnras, 513, 4587, \dodoi{10.1093/mnras/stac769}

\bibitem[{{Weliachew} {et~al.}(1988){Weliachew}, {Casoli}, \& {Combes}}]{Weliachew88}
{Weliachew}, L., {Casoli}, F., \& {Combes}, F. 1988, \aap, 199, 29

\bibitem[{{Whyte} {et~al.}(2002){Whyte}, {Abraham}, {Merrifield}, {Eskridge}, {Frogel}, \& {Pogge}}]{Whyte02}
{Whyte}, L.~F., {Abraham}, R.~G., {Merrifield}, M.~R., {et~al.} 2002, \mnras, 336, 1281, \dodoi{10.1046/j.1365-8711.2002.05879.x}

\end{thebibliography}
\bibliographystyle{aasjournal}


\end{document}